\documentclass[a4paper,11pt]{article}
\usepackage{jheppub} 
\usepackage{lineno}

\usepackage{graphicx}
\usepackage{dcolumn}
\usepackage{bm}

\newcommand{\bq}{\begin{equation}}
\newcommand{\eq}{\end{equation}}
\newcommand{\bqa}{\begin{eqnarray}}
\newcommand{\eqa}{\end{eqnarray}}
\newcommand{\nn}{\nonumber \\}

\def\be     {\begin{equation}}
\def\ee     {\end{equation}}
\def\bea        {\begin{eqnarray}}
\def\eea        {\end{eqnarray}}
\def\bnn    {\begin{eqnarray*}}
\def\enn    {\end{eqnarray*}}

\arxivnumber{~}

\title{Emergent $\text{AdS}_{d+1}$ Geometry from Functional Renormalization Group via Bulk Metric Reconstruction}

\author{Hyeon Jung Kim$^{a}$ and Ki-Seok Kim$^{a,b}$}

\affiliation{$^{a}$Department of Physics, POSTECH, Pohang, Gyeongbuk 37673, Korea \\ $^{b}$Asia Pacific Center for Theoretical Physics (APCTP), Pohang, Gyeongbuk 37673, Korea}
	\emailAdd{hkim7218@postech.ac.kr}
	\emailAdd{tkfkd@postech.ac.kr}

\date{\today}

\abstract{We present a holographic dual description for the O(N) vector model in $d$-dimensional Euclidean space within the functional renormalization group (FRG) framework. 
By iterating Wilsonian renormalization group transformations, the extra-dimensional scale coordinate is identified as the radial direction of an emergent $(d+1)$-dimensional bulk spacetime. 
We construct a bidirectional holographic dictionary in which the emergent bulk metric is determined by the spectrum of the non-perturbative $GG$ polarization function. Using this dictionary we show that as the mass gap decreases, the emergent geometry evolves from a Ricci flat to an Anti-de Sitter $(AdS_{d+1})$ geometry. }

\begin{document}
\maketitle 
\flushbottom
\section{Introduction}

The holographic duality, or the Anti-de Sitter/Conformal Field Theory (AdS$_{d+1}$/CFT$_d$) correspondence, maps strongly coupled boundary quantum field theories to dual gravitational dynamics in a higher-dimensional bulk spacetime \cite{Maldacena:1997re,Gubser:1998bc,Witten:1998qj,Aharony:1999ti}. 
Within this framework, relations between radial bulk evolution and boundary renormalization group (RG) flow have been studied using Hamilton--Jacobi formulations \cite{deBoer:1999tgo}. 
Approaches based on conformal field theory data have also investigated the emergence of local bulk interactions \cite{Heemskerk:2009pn}. 

The functional renormalization group (FRG) provides a formulation of scale-dependent quantum effective actions \cite{Wetterich:1992yh,Morris:1993qb}. 
Radial evolution has also been formulated within holographic Wilsonian RG and holographic renormalization \cite{Faulkner:2010jy,Papadimitriou:2011qb}. 
In formulations where the continuous RG scale is represented as an additional coordinate, the sequence of effective theories can be rewritten as an evolution along an emergent radial direction \cite{RG_Monotonicity_NEQ,RG_Flow_Nonperturbative_String,Kim:2025frg}. 
However, identifying the RG scale with a radial coordinate is not sufficient to determine the bulk geometry. 
A prescription relating scale-dependent field-theoretic quantities, including running couplings and collective fields, to bulk geometric variables is additionally required.

\noindent Several approaches have addressed this problem within exact RG formulations. Motivated by the proposed duality between the singlet sector of the large-$N$ $O(N)$ vector model and higher-spin gauge theory in AdS \cite{Klebanov:2002ja}, exact RG formulations have recast the RG evolution of bilocal sources as radial evolution of higher-spin bulk fields on a prescribed AdS background \cite{Sathiapalan:2014mna}. Related holographic Wilsonian RG analyses developed the correspondence between radial evolution in the bulk and Wilsonian flow in the boundary theory, emphasizing the role of single- and multi-trace operators \cite{Heemskerk:2010hk}, while subsequent studies extended exact RG constructions to scalar sectors and bulk locality \cite{Sathiapalan:2020pzv,Sathiapalan:2023fuh}. These approaches describe the RG evolution of bulk fields on a prescribed background geometry, while the background metric itself remains fixed by the chosen holographic construction.

\noindent A complementary direction is provided by collective-field formulations. Bulk descriptions have been constructed from large-$N$ theories by reformulating the boundary dynamics in terms of bilocal collective fields \cite{Das:2003vw,Jevicki:2023fzp}. Although these approaches successfully establish bulk locality and higher-dimensional dynamics, the background geometry is typically specified within the adopted collective-field framework. As a result, determining the continuous evolution of the bulk metric and curvature directly from non-perturbative RG trajectories remains an open problem.

\noindent Building on these developments, the quantum renormalization group (QRG), which provides a constructive framework in which iterating RG transformations generates an emergent radial coordinate and promotes collective degrees of freedom to bulk fields \cite{Lee:2013dua}. For the large-$N$ $U(N)$ vector model, the matter fields are integrated out, leaving an effective bulk description in terms of scale-dependent bilocal hopping fields. The bulk metric is then reconstructed from the range of the saddle-point bilocal field and by matching its fluctuation equation to a diffusion equation in curved space \cite{SungSik_Holography_IV}. This construction establishes a geometric interpretation of the correlation structure encoded in the collective-field background, although the resulting geometry is restricted to the leading large-$N$ saddle-point approximation.

Tensor-network and quantum-information approaches provide another framework for describing holographic geometry. 
The Multi-scale Entanglement Renormalization Ansatz (MERA) relates the organization of entanglement across scales to a discrete hyperbolic geometry, while holographic quantum error-correcting codes describe how bulk information is encoded in and reconstructed from boundary degrees of freedom \cite{Swingle:2009bg,Almheiri:2014lwa,Pastawski:2015qua,Hayden:2016cgc}. 
These constructions capture aspects of holographic entanglement entropy, bulk encoding, and bulk reconstruction \cite{Ryu:2006bv,Bao:2015uaa}. 
Ensemble-based approaches have also derived Ryu--Takayanagi-type relations and bulk geometries from large-$c$ conformal field theory data in multi-interval and multi-boundary settings \cite{HaoGeng1:2025,HaoGeng2:2025}. 
Kinematic-space and integral-geometric approaches reconstruct classes of static bulk spatial geometries from boundary entanglement data \cite{Czech:2015xna}, while related work examines the role of local symmetries in the kinematic limit of emergent spacetime \cite{Bao:2017guc}. 
The common principle underlying these approaches is that the structure of boundary entanglement and information flow constrains the connectivity and spatial organization of the emergent bulk. 
However, their primary input is entanglement or information-theoretic data rather than the local evolution of running couplings and correlation functions, making it difficult to determine the continuous bulk metric and curvature directly from a dynamical RG flow.

In this work, we develop a continuous holographic dual effective field theory for the $O(N)$ vector model within the FRG framework \cite{Kim:2025frg}. 
Within this formulation, the holographic dictionary is derived directly from the non-perturbative FRG equations without introducing an {\it ad hoc} bulk metric ansatz or kinematic truncation. 
By matching the linearized fluctuation equation obtained from the FRG with the Laplace--Beltrami equation in a curved background, we determine the emergent bulk metric. 
This matching identifies the transverse spatial warp factor as the conformal component of the metric, $g_{xx}(z)$, while the radial metric component, $g_{zz}(z)$, is determined by the non-perturbative matter fluctuations. 
Using a saddle-point analysis governed by an inverse error-function profile, we determine the radial evolution of the bulk metric and its curvature. 
As the mass gap decreases, the emergent geometry continuously evolves toward a pure AdS$_{d+1}$ space satisfying the Einstein-space condition, $R=-\frac{d(d+1)}{l_{AdS}^2}$. 

The remainder of this paper is organized as follows. 
In Section \ref{section2}, the holographic effective field theory is derived via the continuous functional RG framework, and the holographic dictionary is established. 
Section \ref{section3} investigates the physics of the emergent space, focusing on the bulk equations of motion, the semi-analytic inverse error function solution, the verification of local energy conditions, and the $\text{AdS}_{d+1}$ vacuum geometry. 
Finally, Section \ref{section4} summarizes the findings and discusses directions for future research.

\section{A holographic dual effective field theory in the functional RG analysis}\label{section2}

We begin by considering the $O(N)$ vector model in $d$-dimensional Euclidean space:
\bqa && \resizebox{0.85\linewidth}{!}{$\displaystyle Z=\int D\phi_\alpha(x)\exp\left[-\int d^dx\left\{\sum_{\alpha=1}^{N}\left(\partial_\mu\phi_\alpha(x)\right)^2+m^2\sum_{\alpha=1}^{N}\phi_\alpha^2(x)+\frac{u}{2N}\sum_{\alpha=1}^{N}\phi_\alpha^2(x)\sum_{\beta=1}^{N}\phi_\beta^2(x)\right\}\right]$}, 
\label{O(N)_Vector_Model} \eqa
where $\phi_\alpha(x)$ denotes a real scalar field with $N$ components. 
The quartic interaction term is scaled by $1/N$ for the large-$N$ limit. 
To decouple the quartic coupling, we perform a Hubbard-Stratonovich transformation by introducing an auxiliary field $\varphi(x)$:
\bqa && \resizebox{0.85\linewidth}{!}{$\displaystyle Z = \int D \phi_{\alpha}(x) D \varphi(x) \exp\Big[ - \int d^{d} x \Big\{ \phi_{\alpha}(x) \Big( - \partial_{\mu}^{2} + m^{2} - i \varphi(x) \Big) \phi_{\alpha}(x) + \frac{N}{2 u} \varphi^{2}(x) \Big\} \Big] $}. \eqa
In this representation, the field $\varphi(x)$ mediates the interaction between the scalar fields, allowing a saddle-point analysis in the large-$N$ limit.

To investigate the scale-dependent behavior of the theory, we implement the Wilsonian RG transformation. 
The matter field $\phi_\alpha(x)$ is decomposed into low-energy and high-energy modes. 
By performing a Gaussian integration over the high-energy fluctuations within an energy window scaled by the RG interval $dz$, we obtain the following effective action for the low-energy modes:
\bqa && Z = \int D \phi_{\alpha}(x) D \varphi(x) \exp\Big[ - \int d^{d} x \Big\{ \phi_{\alpha}(x) \Big( - \partial_{\mu}^{2} + m^{2} - i \varphi(x) \Big) \phi_{\alpha}(x) + \frac{N}{2 u} \varphi^{2}(x) \nn && + 2 \alpha d z \frac{N}{4} \ln \Big( - \partial_{\mu}^{2} + m^{2} - i \varphi(x) \Big) \Big\} \Big] . \eqa
Here, the parameter $\alpha$ controls the scale step of the RG transformation. 
The trace-log contribution originates from the determinant of the integrated high-energy modes and the Jacobian factor of the field transformation. 
This term represents the non-perturbative corrections to the effective potential of the field $\varphi(x)$ induced by matter fluctuations. 
The derivation of this procedure is provided in Appendix~\ref{appendixA}.

Next, we address the scale-dependent evolution of the collective sector. 
The field $\varphi(x)$ is replaced with an initial configuration $\varphi^{(0)}(x)$, and its corresponding fluctuations are introduced. 
Integrating out the high-energy collective modes and shifting the field via $\varphi^{(1)}(x) \rightarrow \varphi^{(1)}(x) - \varphi^{(0)}(x)$ yield:
\begin{small}
\bqa && Z = \int D \phi_{\alpha}(x) D \varphi^{(1)}(x) D \varphi^{(0)}(x) \exp\Big[ - \int d^{d} x \Big\{ \phi_{\alpha}(x) \Big( - \partial_{\mu}^{2} + m^{2} - i \varphi^{(1)}(x) \Big) \phi_{\alpha}(x) + \frac{N}{2 u} \varphi^{(0) 2}(x) \nn && + \frac{1}{2 \beta d z} \frac{N}{2} \Big( \varphi^{(1)}(x) - \varphi^{(0)}(x) \Big) \Big\{ \frac{1}{u} + \alpha \beta (d z)^{2} \Big( - \partial_{\mu}^{2} + m^{2} - i \varphi^{(0)}(x) \Big)^{-2} \Big\} \Big( \varphi^{(1)}(x) - \varphi^{(0)}(x) \Big) \nn && - 2 \alpha d z \frac{N}{4} \Big( - \partial_{\mu}^{2} + m^{2} - i \varphi^{(0)}(x) \Big)^{-1} i \Big( \varphi^{(1)}(x) - \varphi^{(0)}(x) \Big) + 2 \alpha d z \frac{N}{4} \ln \Big( - \partial_{\mu}^{2} + m^{2} - i \varphi^{(0)}(x) \Big) \Big\} \Big] , \qquad\hspace{0.5em}\eqa
\end{small}
where $\beta$ serves as the control parameter for the RG flow of the collective field. 
The quadratic term in $\left(\varphi^{(1)} - \varphi^{(0)}\right)$ contains the inverse of the Random Phase Approximation (RPA) propagator, which includes the one-loop level RG corrections \cite{Hertz:1976zz,Millis:1993zz,Metzner:2011zz}. 
This recursive structure defines the step-by-step evolution for the continuous bulk dynamics.

To construct a continuous representation along the RG direction, we iterate these transformations. 
In taking the continuum limit, the discrete steps scale based on their power series expansion in $dz$. 
The terms linear in $dz$ form a Riemann integration over the emergent extra dimension $z$, which defines the bulk action \cite{Kim:2025frg, Lee:2013dua, SungSik_Holography_IV, SungSik_Holography_III, Brute_Force_RG_Derivation_Lattice_Kim, Kitaev_Entanglement_Entropy_Kim, Kondo_Holography, RG_GR_Geometry_I_Kim, RG_GR_Geometry_II_Kim, Brute_Force_RG_Derivation_Dirac_Kim, RG_Flow_Direct_Calculation, Nonperturbative_Wilson_RG, Nonperturbative_Wilson_RG_Disorder, Nonperturbative_RG_Flow}. 
Conversely, the higher-order contributions, such as $(dz)^2$ arising from the loop-level vacuum fluctuations, remain finite and define the scale of the RG evolution; we parameterize these higher-order intervals by a regularizing cutoff scale $\epsilon$ (i.e., $dz \rightarrow \epsilon$). 
Through this classification of scale orders, the sequential chain of field evolutions is expressed as a path integral over the RG trajectory:
\bqa && Z = \int D \phi_{\alpha}(x) D \varphi(x,z) \exp\Bigg[ - \int d^{d} x \Bigg\{ \phi_{\alpha}(x) \Big( - \partial_{\mu}^{2} + m^{2} - i \varphi(x,z_{f}) \Big) \phi_{\alpha}(x) + \frac{N}{2 u} \varphi^{2}(x,0) \Bigg\} \nn && - N \int_{0}^{z_{f}} d z \int d^{d} x \Bigg\{ \frac{1}{4 \beta} \Big( \partial_{z} \varphi(x,z) \Big) \Bigg( \frac{1}{u} + \alpha \beta \varepsilon^{2} \Big( - \partial_{\mu}^{2} + m^{2} - i \varphi(x,z) \Big)^{-2} \Bigg) \Big( \partial_{z} \varphi(x,z) \Big) \nn && - i \alpha \frac{\varepsilon}{2} \Big( - \partial_{\mu}^{2} + m^{2} - i \varphi(x,z) \Big)^{-1} \Big( \partial_{z} \varphi(x,z) \Big) + \frac{\alpha}{2} \ln \Big( - \partial_{\mu}^{2} + m^{2} - i \varphi(x,z) \Big) \Bigg\} \Bigg] .\label{eq:bulk_action} \eqa

Based on this derivation, the holographic dual effective field theory is defined as:
\bqa && \begin{split}Z = \int D \phi_{\alpha}(x) D \varphi(x,z) \exp\Big[ &- \int d^{d} x \Big\{ \phi_{\alpha}(x) G^{-1}(x,z_{f}) \phi_{\alpha}(x) + \frac{N}{2 u} \varphi^{2}(x,0) \Big\} \\
&- N \int_{0}^{z_{f}} d z \int d^{d} x ~ \mathcal{L}_{bulk} \Big] ,\end{split} \eqa
where the bulk Lagrangian density $\mathcal{L}_{\text{bulk}}$ is given by:
\begin{small}\bqa && \mathcal{L}_{bulk}=\frac{1}{4\beta }\left(\partial _{z}\varphi (x,z)\right)D^{-1}(x,z)\left(\partial _{z}\varphi (x,z)\right)-i\alpha \frac{\epsilon }{2}G(x,z)\left(\partial _{z}\varphi (x,z)\right)+\frac{\alpha }{2}\ln G^{-1}(x,z) . \label{Bulk_Lagrangian} \eqa\end{small}
The scale-dependent Green's functions $G(x, z)$ and $D(x, z)$ are related to the collective field configuration via:
\bqa && \Big( - \partial_{\mu}^{2} + m^{2} - i \varphi(x,z) \Big) G(x,z) = \delta^{(d)}(x) , \label{GF_G} \\ && \Big( \frac{1}{u} + \alpha \beta \varepsilon^{2} G^{2}(x,z) \Big) D(x,z) = \delta^{(d)}(x) . \label{GF_D} \eqa
The propagators $G$ and $D$ define the background of the $(d+1)$-dimensional theory, where the radial boundaries $z = 0$ and $z = z_f$ correspond to the UV and IR regimes of the boundary QFT, respectively. For a more detailed discussion on its direct connection to the FRG framework, the reader is referred to Appendix~\ref{appendixC}.

\section{Emergent $\text{AdS}_{d+1}$ geometry}\label{section3}

\subsection{Bulk equation and UV $\&$ IR boundary conditions}

The classical dynamics of the field $\varphi(x, z)$ within the emergent $(d + 1)$-dimensional bulk is determined by the principle of least action. 
Taking the functional variation of the bulk action in Eq. (\ref{Bulk_Lagrangian}) with respect to $\varphi(x, z)$ yields the following equation of motion:
\bqa && - \partial_{z} \Big\{ D^{-1}(x,z) \Big( \partial_{z} \varphi(x,z) \Big) \Big\} = i \alpha \beta G(x,z) . \label{Bulk_Eq} \eqa
This equation describes the evolution of the field along the radial direction $z$. 
The left-hand side, which contains the propagator $D(x, z)$, represents the kinetic evolution of the interactions across energy slices. 
The right-hand side is proportional to the matter-field propagator $G(x, z)$, which acts as a source term that represents the feedback of microscopic fluctuations. 
In the framework of holographic duality, Eq. (\ref{Bulk_Eq}) expresses the RG flow of the boundary theory as a bulk field equation from the UV boundary ($z = 0$) toward the IR regime ($z = z_f$).

To obtain a solution for this field configuration, we specify boundary conditions at both extremities of the holographic dimension \cite{Kim:2025frg,Lee:2013dua, SungSik_Holography_IV, SungSik_Holography_III,  Brute_Force_RG_Derivation_Lattice_Kim, Kitaev_Entanglement_Entropy_Kim, Kondo_Holography, RG_GR_Geometry_I_Kim, RG_GR_Geometry_II_Kim, Brute_Force_RG_Derivation_Dirac_Kim, RG_Flow_Direct_Calculation, Nonperturbative_Wilson_RG, Nonperturbative_Wilson_RG_Disorder, Nonperturbative_RG_Flow}. 
These conditions require the effective action to be stationary at the boundaries. 
Based on the variational analysis and the definition of the canonical momentum $\Pi_\varphi(x, z)$ in Appendix~\ref{appendixB}, the field variation at the IR boundary ($z = z_f$) enforces:
\bqa && \partial_{z_{f}} \varphi(x,z_{f}) = i \beta (1 + \varepsilon\alpha) D(x,z_{f}) G(x,z_{f}) . \label{IR_BC} \eqa 
This IR constraint relates to the boundary conditions imposed by the matter-field fluctuations at the infrared scale. 
Simultaneously, the variation at the UV boundary ($z = 0$) requires the field to satisfy:
\bqa && [\partial_{z} \varphi(x,z)]_{z = 0} = \frac{2 \beta}{u} D(x,0) \varphi(x,0) + i \varepsilon \alpha\beta D(x,0) G(x,0) . \label{UV_BC} \eqa
Equation (\ref{UV_BC}) defines the UV boundary condition, relating the bulk dynamics to the parameters, such as the quartic coupling constant $u$, of the boundary field theory. 
The bulk differential equation and these boundary conditions form a boundary value problem for the emergent system.

\subsection{Approximate solution}

We investigate the spatial and radial profiles of the field $\varphi(x, z)$ by solving the non-linear equation. 
Since the propagators $G(x, z)$ and $D(x, z)$ depend on the transverse coordinates $x$ through the relations in Eqs. (\ref{GF_G}) and (\ref{GF_D}), the differential system captures the inhomogeneous geometric structure.

\noindent To analysis this system semi-analytically, we consider the regime in which the RPA corrections are governed by the coupling $u$. 
In this limit, the bulk equation of motion (Eq.~(\ref{Bulk_Eq})) reduces to a second-order differential equation:
\bqa && - \partial_{z}^{2} \varphi(x,z) = i \alpha \beta u G(x,z) . \label{Bulk_Eq_Simple} \eqa
In this representation, the evolution of the field along the radial coordinate $z$ is driven by the matter propagator $G(x, z)$. 
The boundary conditions in Eqs. (\ref{IR_BC}) and (\ref{UV_BC}) are evaluated under this same limit, yielding the following boundary criteria:
\bqa && \partial_{z_{f}} \varphi(x,z_{f}) = i \beta u G(x,z_{f}) \label{IR_BC_Simple} , \\ && [\partial_{z} \varphi(x,z)]_{z = 0} = 2 \beta \varphi(x,0) . \label{UV_BC_Simple} \eqa
Here, Eq.~(\ref{IR_BC_Simple}) specifies the derivative at the infrared boundary $z_{f}$, representing the feedback from the matter sector at each spatial coordinate $x$. 
Meanwhile, Eq.~(\ref{UV_BC_Simple}) sets the matching condition at the UV boundary $z = 0$, linking the bulk scaling behavior to the boundary value of the field. 
These expressions (Eqs.~(\ref{Bulk_Eq_Simple})--(\ref{UV_BC_Simple})) form a closed boundary value problem used to determine the analytical form of the emergent geometry.

\begin{figure}[h]
\centering
\begin{tabular}{cc}
\includegraphics[width=0.47\textwidth]{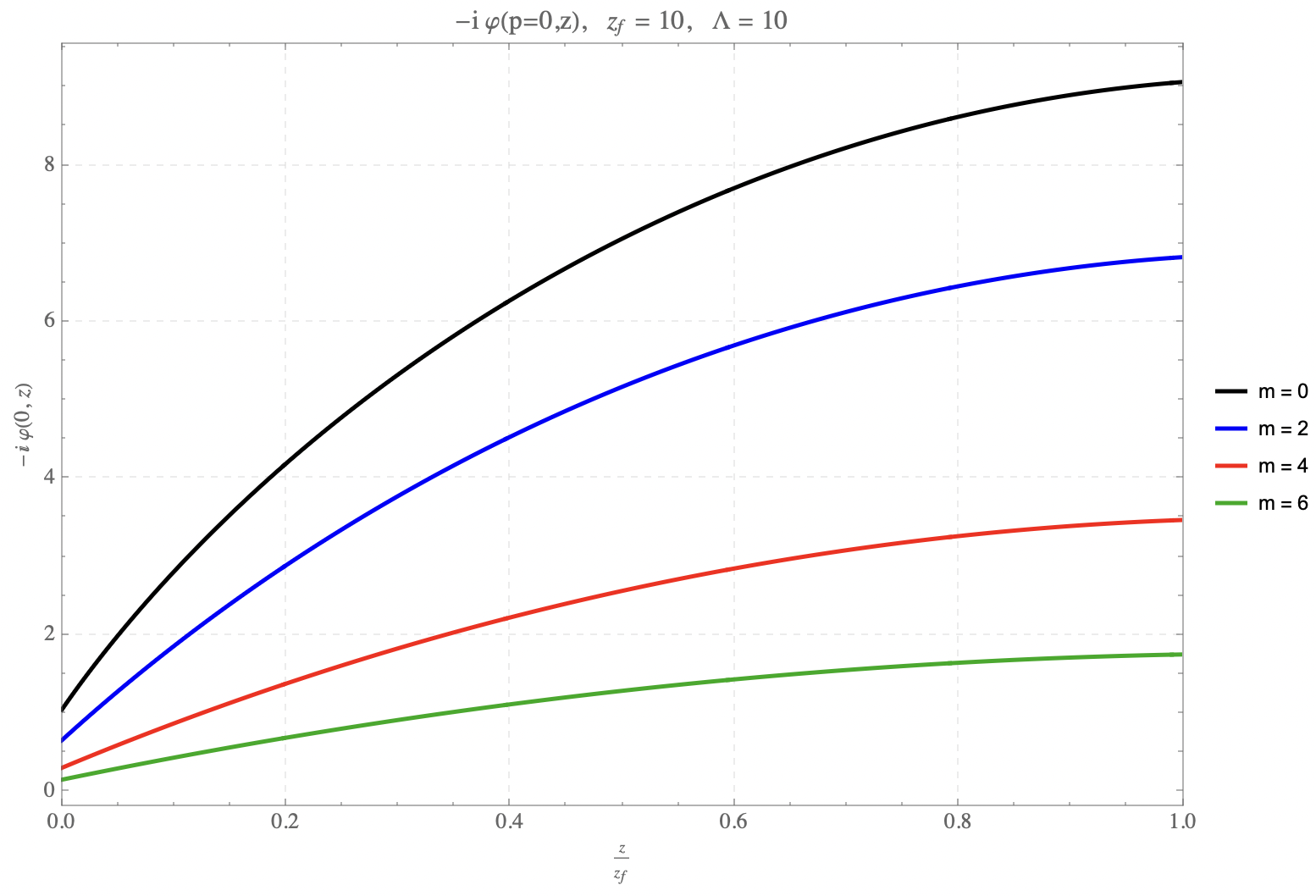}&\includegraphics[width=0.47\textwidth]{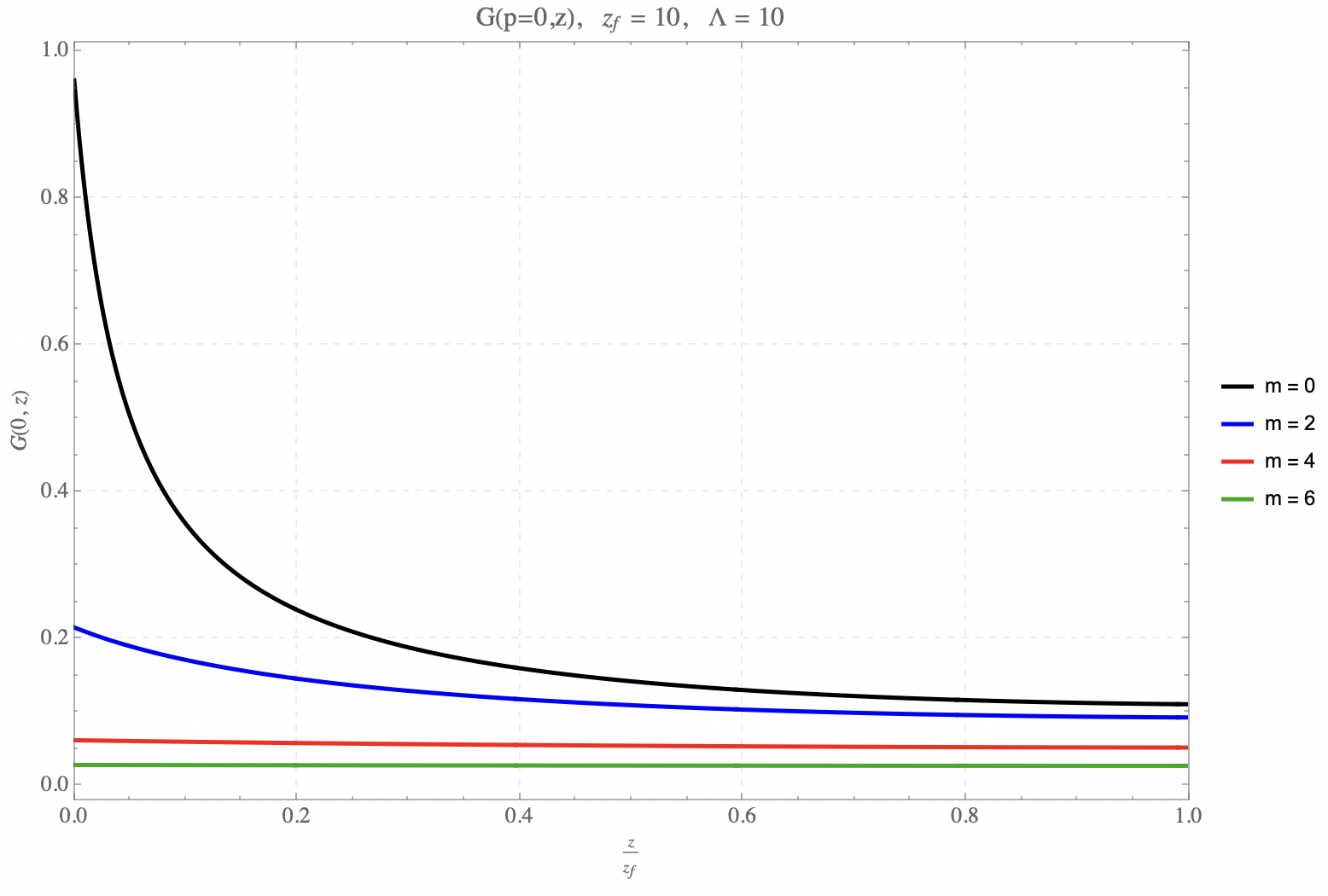}\\
\text{(a) }$\varphi(p=0,z)$ &\text{(b) }$G(p=0,z)$
\end{tabular}
\caption{Numeric calculation of the collective mode $\varphi$ and Green function $G$ for $m=0, 2, 4$, and $6$ under the parameters $\alpha = \beta = u = 1$, and $z_f=10$.}\label{fig:green_function_profiles}
\end{figure}

Integrating the second-order bulk differential equation along the radial dimension yields the following profile: 
\bqa && \varphi(p,z) = i \Bigg( e^{- \frac{\mathcal{C}_{1}(p)}{2 \alpha \beta u}} e^{- \Big\{ InverseErf\Big[ \sqrt{\frac{2 \alpha \beta u}{\pi}} e^{\frac{\mathcal{C}_{1}(p)}{2 \alpha \beta u}} \Big(z + \mathcal{C}_{2}(p)\Big)\Big] \Big\}^{2}} - \Big(p^2+m^2\Big) \Bigg) , \label{Solution} \qquad\hspace{0.5em}\eqa
where, $\mathcal{C}_1(p)$ and $\mathcal{C}_2(p)$ are integration functions fixed by the UV and IR boundary conditions. Introducing
\bqa && R_0(p)\equiv InverseErf\Big[\sqrt{\frac{2\alpha\beta u}{\pi}e^{\frac{C_1(p)}{\alpha\beta u}}}\mathcal{C}_2(p)\Big], \eqa 
\bqa && R_f(p)\equiv InverseErf\Big[\sqrt{\frac{2\alpha\beta u}{\pi}e^{\frac{C_1(p)}{\alpha\beta u}}}\Big(z_f+\mathcal{C}_2(p)\Big)\Big]. \eqa
The integration functions are expressed as
\bqa && \mathcal{C}_{1}(p) = -2\alpha\beta u \ln \bigg[-\frac{\beta u}{\sqrt{2\alpha\beta u}R_f(p)}e^{R_f^2(p)}\bigg] , \label{C1_Function}\eqa
\bqa && \mathcal{C}_{2}(p) = -\frac{\sqrt{\pi}}{2\alpha}\frac{e^{R_f^2(p)}}{R_f(p)}Erf[R_0(p)] ,\label{C2_Function} \eqa
where, $R_0(p)$ and $R_f(p)$ satisfy
\bqa && \operatorname{Erf}[R_f(p)]-\operatorname{Erf}[R_0(p)]=-\frac{2\alpha z_f}{\sqrt{\pi}}R_f(p)e^{-R_f(p)^2} , \label{eq:boundary_condition1} \eqa
\bqa && -\sqrt{2\alpha\beta u}\,R_0(p)=2\beta\left[-\frac{\beta u}{\sqrt{2\alpha\beta u}\,R_f(p)}e^{R_f(p)^2-R_0(p)^2}-(p^2+m^2)\right]. \label{eq:boundary_condition2} \eqa
The Inverse Error Function appearing in Eq.~(\ref{Solution}) governs the radial evolution of the collective field. Its mathematical properties and asymptotic expansions are summarized in Refs.~\cite{Abramowitz:1972,Olver:2010}. The corresponding matter propagator is obtained as
\begin{small}\bqa && G(p,z)=-\frac{\sqrt{2\alpha\beta u}}{\beta u}R_f(p)e^{-R_f(p)^2}\exp{\bigg[InverseErf\Big(\Big(1-\frac{z}{z_f}\Big)Erf R_0(p)+\frac{z}{z_f}Erf R_f(p)\Big)\bigg]^2}.\nn&& \label{Solution_Green_Function} \eqa\end{small}
The analytic solutions in Eqs.~(\ref{Solution}) and (\ref{Solution_Green_Function}) describe the nonlinear evolution of the collective field and the corresponding Green's function along the holographic direction. These solutions provide the starting point for determining the emergent bulk metric.
Fig.~\ref{fig:green_function_profiles} presents the numerically calculated profiles of the collective field $\varphi(p=0,z)$ and the Green's function $G(p=0,z)$ for several values of the bare mass parameter $m$.

\subsection{Normal modes} 

\subsubsection{Linearized equation of motion} 
To investigate the excitations in the bulk, we consider the fluctuations of the field around its saddle-point configuration, defined as $\varphi(x, z) = \varphi_c(x, z) + \delta\varphi(x, z)$. 
Linearizing the bulk equation of motion (Eq. (\ref{Bulk_Eq})) with respect to the fluctuation $\delta\varphi(x, z)$ yields the following evolution equation: 
\bqa && - \partial_{z} \Big\{ D^{-1}(x,z) \Big( \partial_{z} \delta \varphi(x,z) \Big) \Big\} = - \alpha \beta G^{2} (x,z) \delta \varphi(x,z) . \label{Normal_Mode_Bulk} \eqa
In this framework, the background Green's functions $G(x, z)$ and $D(x, z)$ are evaluated at the saddle-point configuration $\varphi_c(x, z)$, satisfying the relations:
\bqa && \Big( - \partial_{\mu}^{2} + m^{2} - i \varphi_{c}(x,z) \Big) G(x,z) = \delta^{(d)}(x) \label{Normal_Mode_IR_BC} , \\ && \Big( \frac{1}{u} + \alpha \beta \varepsilon^{2} G^{2}(x,z) \Big) D(x,z) = \delta^{(d)}(x) \label{Normal_Mode_UV_BC} . \eqa
Equation (\ref{Normal_Mode_Bulk}) describes the propagation of the fluctuations within the $(d+1)$-dimensional holographic space. 
The term on the right-hand side, proportional to $G^2(x, z)$, acts as a spatially modulated mass-squared term or potential barrier for the fluctuations along the RG direction. 
Since $G(x, z)$ contains the spatial dependence of the background, the resulting dynamics of $\delta\varphi(x, z)$ are inhomogeneous. 
This linearized structure is used to identify the emergent gravitational metric by mapping the propagation of these fluctuations to the behavior of a scalar field in a curved $(d+1)$-dimensional geometry.

\subsubsection{Emergent geometry}

The inhomogeneous dynamics of the fluctuations $\delta\varphi(x, z)$ allows the RG flow of the $O(N)$ vector model to be interpreted within a curved $(d+1)$-dimensional spacetime. 
We introduce a diagonal metric ansatz that accounts for spatial and radial variations:
\bqa && d s^{2} = g_{zz} d z^{2} + g_{\mu\nu} d x^{\mu} d x^{\nu} . \label{Diagonal_Metric} \eqa
In this geometric framework, the linearized equation of motion for the fluctuations $\delta\varphi(x, z)$ takes the form of a Laplace-Beltrami equation within this curved background \cite{Witten:1998qj,Jost:2008,Polchinski:1998rq}:
\bqa && \frac{1}{\sqrt{g}} \partial_{z} \Big( \sqrt{g} g^{zz} \partial_{z} \delta \varphi(x,z) \Big) + \frac{1}{\sqrt{g}} \partial_{\mu} \Big( \sqrt{g} g^{\mu\nu} \partial_{\nu} \delta \varphi(x,z) \Big) = m_{eff}^{2}(x,z) \delta \varphi(x,z) . \label{KG_General} \eqa
To establish a mapping between the field-theoretic variables and the geometric metric components, the linearized field equation (Eq.~(\ref{Normal_Mode_Bulk})) is rearranged to match the structure of the differential operators in Eq.~(\ref{KG_General}). 
Although the first bulk term in the bulk action (Eq.~(\ref{eq:bulk_action})) explicitly contains only radial derivatives, By iterating Wilsonian renormalization group transformations, logarithmic contributions are expected to be generated in the effective action. Expanding the resulting logarithmic term around the background field may then induce the leading transverse second-derivative term, $-\partial_\mu^2\delta\varphi(x,z)$, in the linearized bulk equation. Accordingly, we include this leading transverse contribution and write the extended equation as
\bqa && \partial_{z} \Big\{ D^{-1}(x,z) \Big( \partial_{z} \delta \varphi(x,z) \Big) \Big\}+\partial_{\mu}^2\delta\varphi(x,z) = \alpha \beta G^2(x,z)\delta \varphi(x,z)  . \label{KG_Our_Case} \eqa
This form enables a comparison between the radial and spatial derivative terms, which determines the metric warping factors and the effective bulk mass through the microscopic propagators.

Matching the coefficients of the differential operators in Eqs. (\ref{KG_General}) and (\ref{KG_Our_Case}) determines the emergent metric in terms of the microscopic propagators:
\bqa && d s^{2} = [D(x,z)]^{\frac{d-2}{d-1}}dz^2+[D(x,z)]^{-\frac{1}{d-1}}\delta_{\mu\nu}dx^\mu dx^\nu. \label{Emergent_Metric} \eqa
This geometry forms a curved background where the RG scale $z$ acts as the radial dimension. 
The mapping also defines the effective bulk mass of the scalar excitations:
\bqa && m_{eff}^{2}(x,z) = \frac{\alpha \beta}{\sqrt{g}} G^2(x,z) . \label{Effective_Mass} \eqa
Equation (\ref{Effective_Mass}) shows that the effective mass in the bulk depends on the bare mass $m^{2}$, the field $\varphi(x,z)$, and the spatial derivatives of the matter-field propagator. 
This geometric dictionary (Eqs. (\ref{Emergent_Metric})--(\ref{Effective_Mass})) relates the non-perturbative RG flow of the $O(N)$ vector model to a dual description of gravity, where the mass of bulk excitations reflects the renormalization of the boundary theory.
\begin{figure}[h]
\centering
\begin{tabular}{cc}
\includegraphics[width=0.5\linewidth]{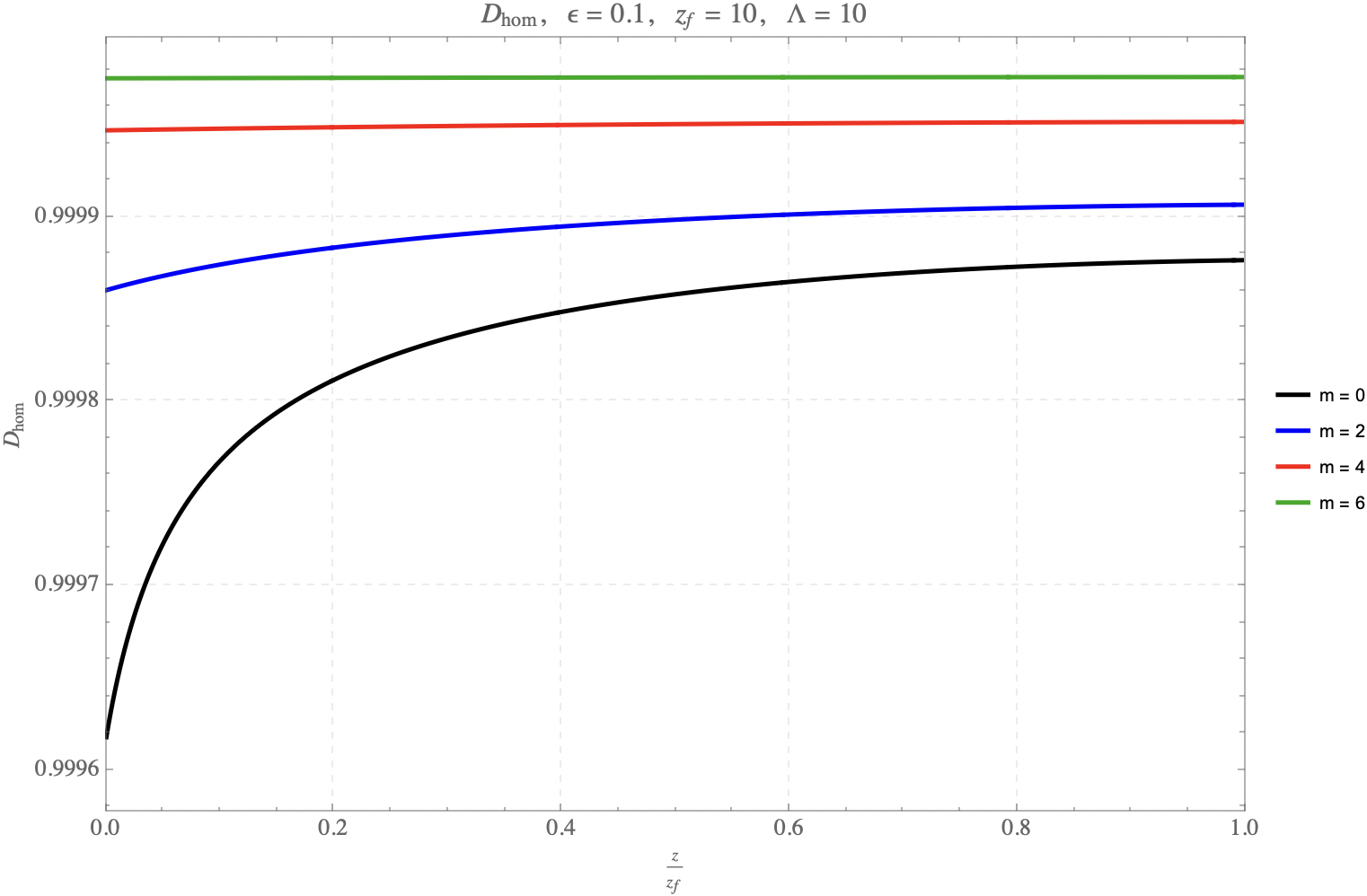}& \includegraphics[width=0.5\linewidth]{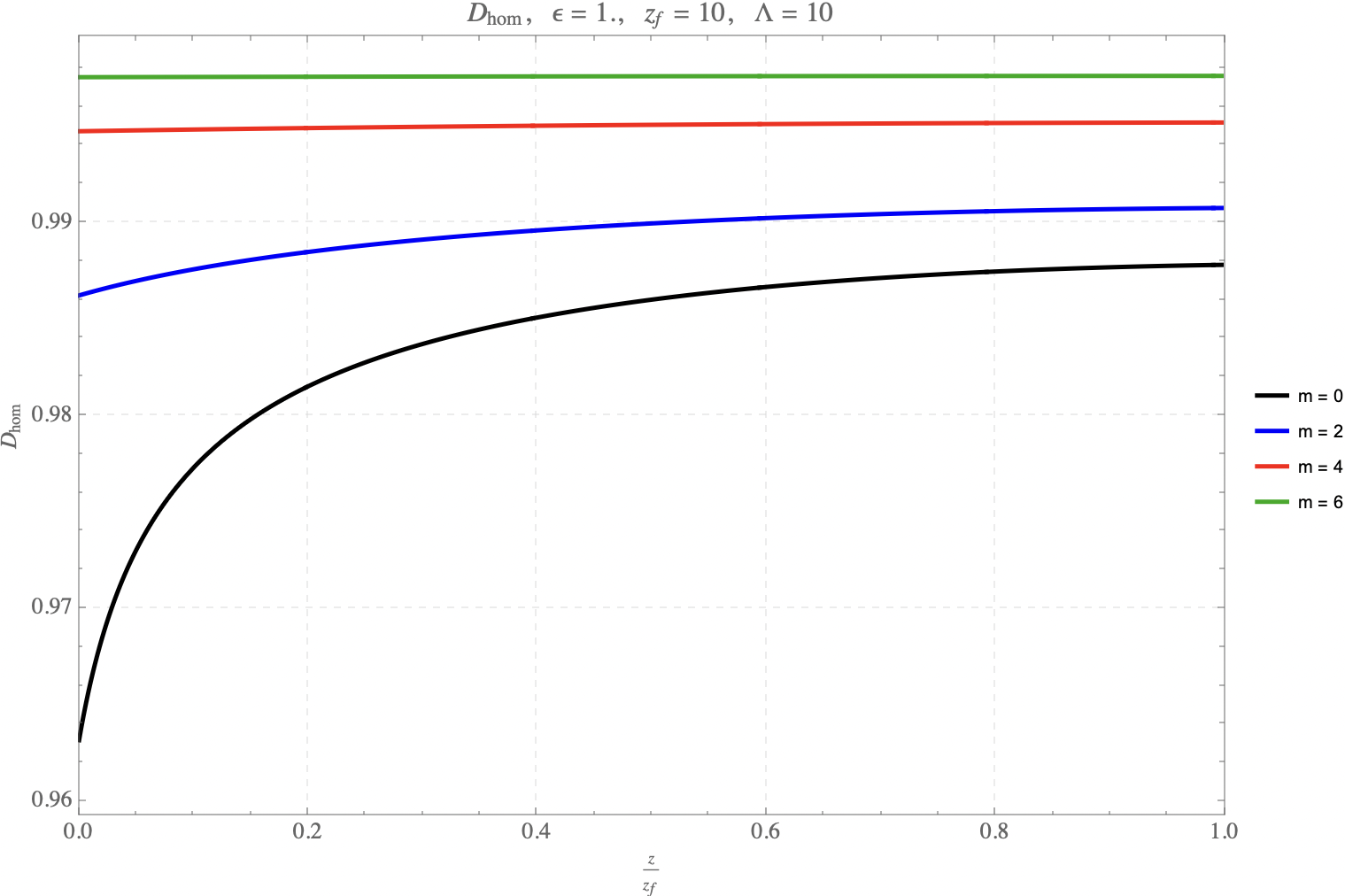}\\
\text{(a) }$\epsilon=0.1$ &\text{(b) }$\epsilon=1.0$
\end{tabular}
\caption{Numeric calculation of the collective mode propagator $D(p=0,z)$ for $m=0, 2, 4$, and $6$ under the parameters $\alpha = \beta = u = 1,$ $d = 3$, and $z_f=10$.}\label{fig:energy_conditions}
\end{figure}
The explicit components of the emergent bulk metric are determined by the local configurations of the propagators $G(x, z)$ and $D(x, z)$. 
Based on the geometric mapping established above, the radial component ($g_{zz}$) and the transverse spatial components ($g_{xx}$) can be determined as:
\bqa && g_{zz}(x,z) =[D(x,z)]^{\frac{d-2}{d-1}} \label{GF_Metric_ZZ} , \\ && g_{xx}(x,z) = [D(x,z)]^{-\frac{1}{d-1}} , ~~~~~ g_{\mu\nu}(x,z) = g_{xx}(x,z) \delta_{\mu\nu} \label{GF_Metric_XX} . \eqa
These relations describe how the collective dynamics of the $O(N)$ vector model relate to the $(d+1)$-dimensional curved background. 

For the diagonal metric Eq. (\ref{Diagonal_Metric}), the non-vanishing components of the Ricci tensor are determined by the radial evolution of the warping factors. 
The radial component, $R_{zz}$, is given by:
\bqa && \begin{split}R_{zz}(x,z) =& -\frac{d}{2}\frac{\partial_z^2 g_{xx}}{g_{xx}}+\frac{d}{4}\frac{(\partial_z g_{xx})^2}{g_{zz}g_{xx}}+\frac{d}{4}\frac{(\partial_z g_{xx})^2}{g_{xx}^2}-\frac{1}{2}\frac{\partial^\rho\partial_\rho g_{zz}}{g_{xx}}+\frac{1}{4}\frac{\partial_\rho g_{zz}\,\partial^\rho g_{zz}}{g_{zz}g_{xx}}\\
&+\frac{2-d}{4}\frac{\partial_\rho g_{zz}\,\partial^\rho g_{xx}}{g_{xx}^2} \label{Ricci_ZZ} . \end{split}\eqa 
This component represents the curvature along the radial direction as the scale changes from the UV to the IR. 
The off diagonal part $R_{z\mu}$:
\bqa && R_{z\mu}=(d-1)\left[-\frac{1}{2}\frac{\partial_z\partial_\mu g_{xx}}{g_{xx}}+\frac{1}{2}\frac{(\partial_z g_{xx})(\partial_\mu g_{xx})}{g_{xx}^{2}}+\frac{1}{4}\frac{(\partial_\mu g_{zz})(\partial_z g_{xx})}{g_{zz}g_{xx}}\right] \label{Ricci_Zmu}. \eqa
Similarly, the spatial components are determined by the coefficient $R_{\mu\nu}$:
\bqa
 \begin{split} 
 R_{\mu\nu}(x,z)=&-\frac{1}{2}\frac{\partial_\mu\partial_\nu g_{zz}}{g_{zz}}-\frac{d-2}{2}\frac{\partial_\mu\partial_\nu g_{xx}}{g_{xx}}+\frac{1}{4}\frac{(\partial_\mu g_{zz})(\partial_\nu g_{zz})}{g_{zz}^{2}}+\frac{3(d-2)}{4}\frac{(\partial_\mu g_{xx})(\partial_\nu g_{xx})}{g_{xx}^{2}}\\
&+\frac{1}{4}\frac{(\partial_\mu g_{zz})(\partial_\nu g_{xx})+(\partial_\nu g_{zz})(\partial_\mu g_{xx})}{g_{zz}g_{xx}}+\delta_{\mu\nu}\bigg[-\frac{1}{2}\frac{\partial_z^2 g_{xx}}{g_{zz}}+\frac{1}{4}\frac{(\partial_z g_{zz})(\partial_z g_{xx})}{g_{zz}^{2}}\\
&+\frac{2-d}{4}\frac{(\partial_z g_{xx})^{2}}{g_{zz}g_{xx}}-\frac{1}{2}\frac{\partial^\rho\partial_\rho g_{xx}}{g_{xx}}-\frac{1}{4}\frac{\partial_\rho g_{zz}\partial^\rho g_{xx}}{g_{zz}g_{xx}}+\frac{4-d}{4}\frac{\partial_\rho g_{xx}\partial^\rho g_{xx}}{g_{xx}^{2}}\bigg] .
 \end{split}
 \label{Ricci_XX} 
 \qquad\hspace{0.2em}
 \eqa
 Contracting the Ricci tensor with the inverse metric yields the Ricci scalar $R$:
\bqa  \begin{split}R(x,z)=&-d\frac{\partial_z^2 g_{xx}}{g_{zz}g_{xx}}+\frac{d}{2}\frac{(\partial_z g_{zz})(\partial_z g_{xx})}{g_{zz}^{2}g_{xx}}+\frac{d(3-d)}{4}\frac{(\partial_z g_{xx})^{2}}{g_{zz}g_{xx}^{2}}-\frac{\partial_\rho\partial^\rho g_{zz}}{g_{zz}g_{xx}}-(d-1)\frac{\partial^\rho\partial_\rho g_{xx}}{g_{xx}^{2}}\\
&+\frac{1}{2}\frac{\partial_\rho g_{zz}\partial^\rho g_{zz}}{g_{zz}^{2}g_{xx}}+\frac{2-d}{2}\frac{\partial_\rho g_{zz}\partial^\rho g_{xx}}{g_{zz}g_{xx}^{2}}-\frac{(d-1)(d-6)}{4}\frac{\partial_\rho g_{xx}\partial^\rho g_{xx}}{g_{xx}^{3}}.\end{split} \label{Ricci_Scalar}  \qquad\hspace{0.1em}\eqa
In Eq. (\ref{Ricci_Scalar}), the Ricci scalar depends on the spatial dimension $d$ of the boundary theory. 
These expressions relate the quantum fluctuations of the $O(N)$ vector model to the curved gravitational background.

To relate the microscopic variables to the emergent gravity, the curvature tensors are expressed in terms of the matter-field propagator $G(x, z)$ and the interaction $D(x, z)$. 
Substituting the holographic dictionary (Eqs. (\ref{GF_Metric_ZZ})--(\ref{GF_Metric_XX})) into the geometric definitions (Eqs. (\ref{Ricci_ZZ})--(\ref{Ricci_XX})) yields the following expressions for the Ricci tensor components:
\bqa && \begin{split}R_{zz}(x,z) =&\frac{d}{2(d-1)}\frac{\partial_z^2D(x,z)}{D(x,z)}-\frac{3d}{4(d-1)}\frac{\left[\partial_zD(x,z)\right]^2}{D(x,z)^2}-\frac{d-2}{2(d-1)}\delta^{\mu\nu}\partial_\mu\partial_\nu D(x,z)\\
&+\frac{d-2}{2(d-1)}\frac{\delta^{\mu\nu}\left[\partial_\mu D(x,z)\right]\left[\partial_\nu D(x,z)\right]}{D(x,z)},\end{split} \label{Ricci_ZZ_GF} \eqa
\bqa && \begin{split}R_{z\mu}(x,z)=\frac{\partial_z\partial_\mu D(x,z)}{2D(x,z)}-\frac{3d-4}{4(d-1)}\frac{\left[\partial_zD(x,z)\right]\left[\partial_\mu D(x,z)\right]}{D(x,z)^2},\end{split} \label{Ricci_muZ_GF} \eqa
\bqa && \begin{split}R_{\mu\nu}(x,z)=&-\frac{d-2}{4(d-1)}\frac{\left[\partial_\mu D(x,z)\right]\left[\partial_\nu D(x,z)\right]}{D(x,z)^2}+\frac{\delta_{\mu\nu}}{2(d-1)}\bigg[\frac{\partial_z^2D(x,z)}{D(x,z)^2}-2\frac{\left[\partial_zD(x,z)\right]^2}{D(x,z)^3}\\
&+\frac{\delta^{\rho\sigma}\partial_\rho\partial_\sigma D(x,z)}{D(x,z)}-\frac{\delta^{\rho\sigma}\left[\partial_\rho D(x,z)\right]\left[\partial_\sigma D(x,z)\right]}{D(x,z)^2}\bigg].
\end{split} \label{Ricci_XX_GF} \qquad\hspace{0.3em}\eqa
%
%
The Ricci scalar is expressed as:
\bqa &&\begin{split} R(x,z)=&\frac{d}{d-1}D(x,z)^{-\frac{2d-3}{d-1}}\partial_z^2D(x,z)-\frac{7d}{4(d-1)}D(x,z)^{-\frac{3d-4}{d-1}}\left[\partial_zD(x,z)\right]^2\\
&+\frac{D(x,z)^{-\frac{d-2}{d-1}}\delta^{\mu\nu}\partial_\mu\partial_\nu D(x,z)}{d-1}-\frac{d+2}{4(d-1)}D(x,z)^{-\frac{2d-3}{d-1}}\delta^{\mu\nu}\left[\partial_\mu D(x,z)\right]\left[\partial_\nu D(x,z)\right]. \end{split}\label{Ricci_Scalar_GF} \eqa
This representation maps the functional RG equations of the $O(N)$ vector model onto the vacuum structure of the emergent curved spacetime. 
%
%
To investgate whether this background geometry AdS or not, we check Ricci scalar R is negative and some critical point $z_c$ it maintain constant value about $z$.
\begin{figure}[ht]
\centering
\begin{tabular}{cc}
\includegraphics[width=0.47\linewidth]{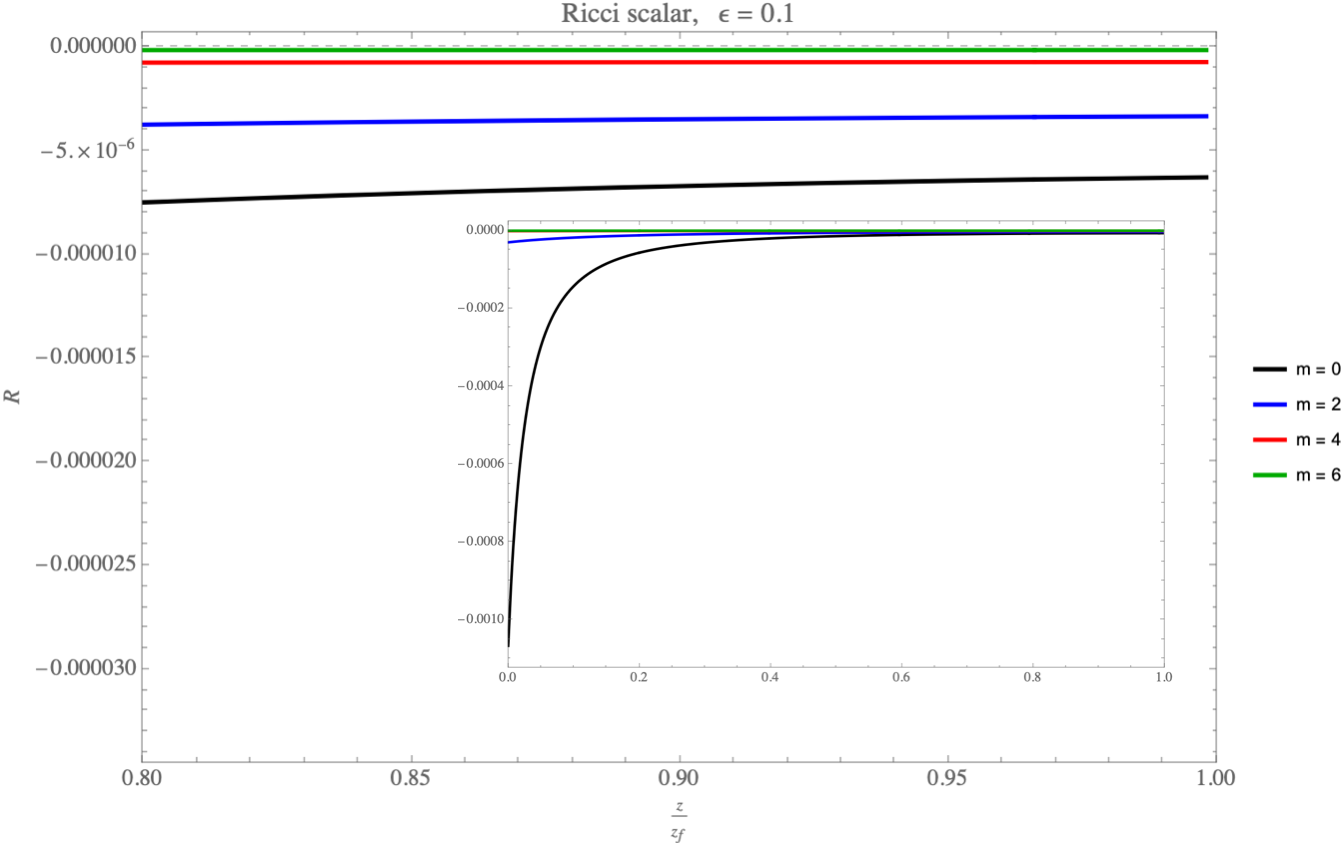}&\includegraphics[width=0.47\linewidth]{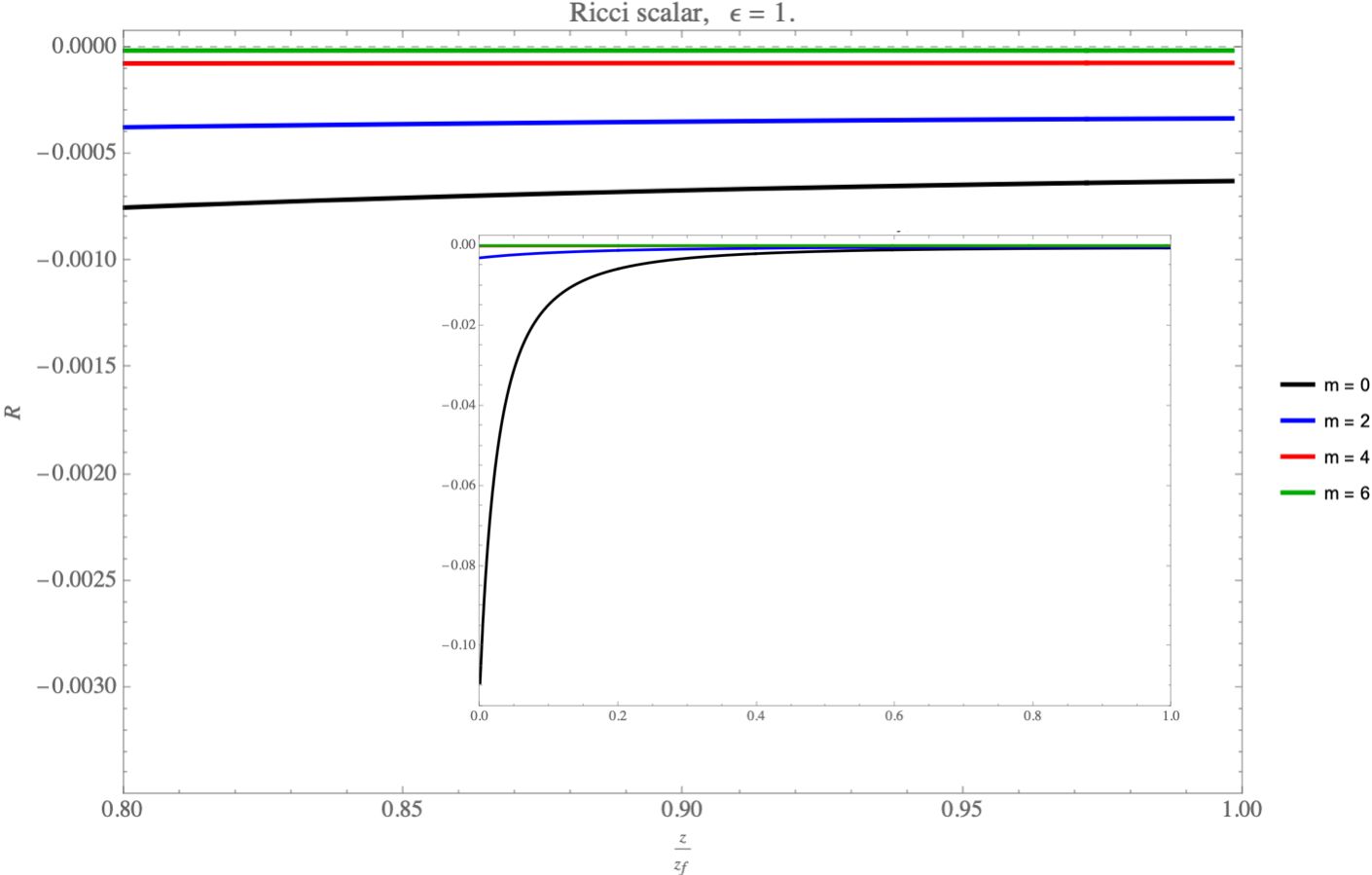}\\
[-0.1cm]\(\text{(a) }\text{Ricci scalar for  }\epsilon=0.1\)&\(\text{(b) }\text{Ricci scalar for }\epsilon=1\)\end{tabular}\caption{Ricci scalar in the long-wavelength limit ($p=0$) at $d=3$, obtained from the first two terms of Eq.~(\ref{Ricci_Scalar_GF}). (a) and (b) correspond to $\epsilon=0.1$ and $\epsilon=1.0$, respectively. In each panel, the Ricci scalar is shown for $0.8 \le z/z_f \le 1.0$, with an inset displaying the full profile over the range $0 \le z/z_f \le 1.0$.}\label{fig:ricci_scalar_profiles}\end{figure}

\begin{figure}[ht]
\centering
\begin{tabular}{cc}
\includegraphics[width=0.47\linewidth]{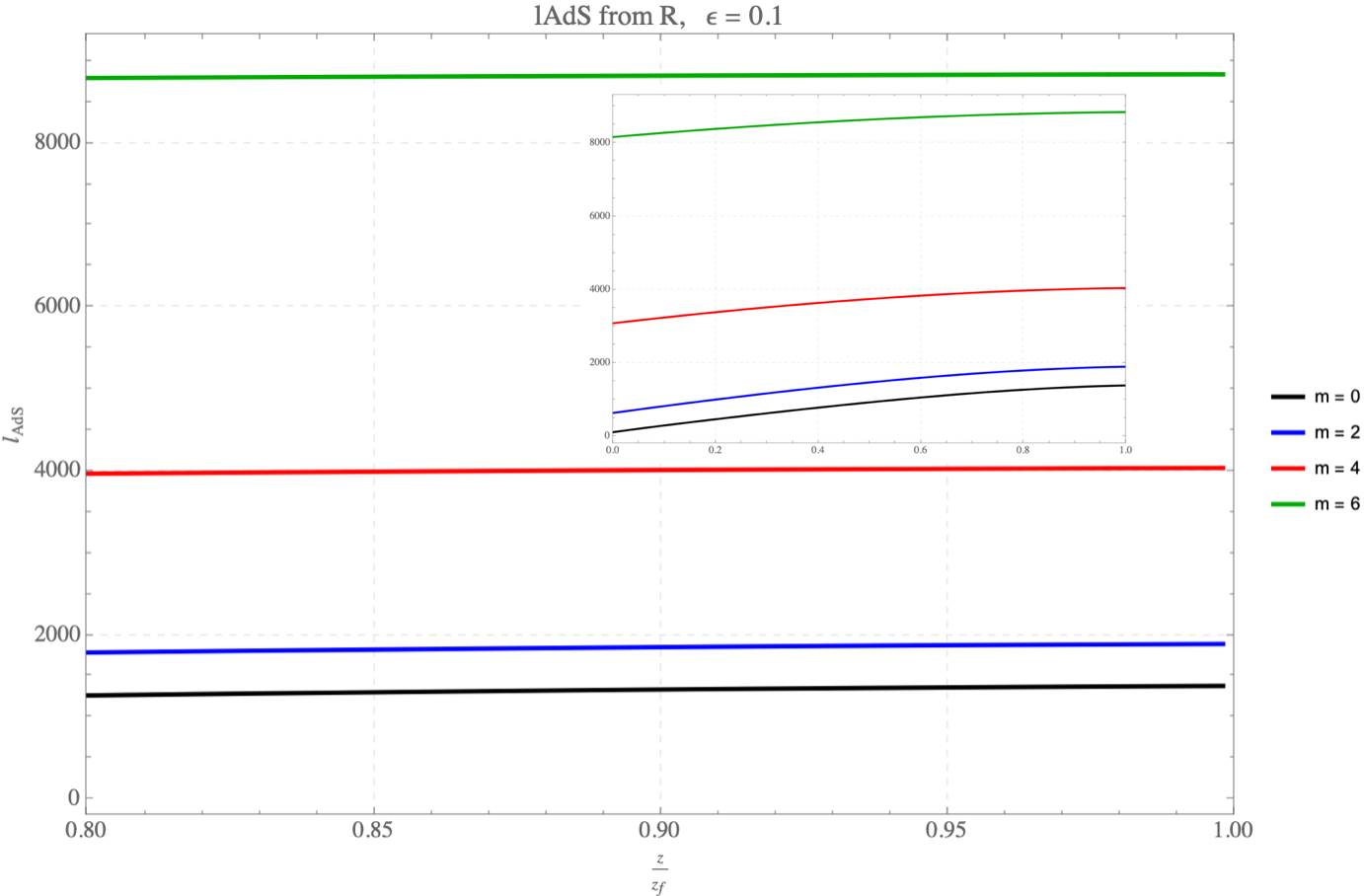}&\includegraphics[width=0.47\linewidth]{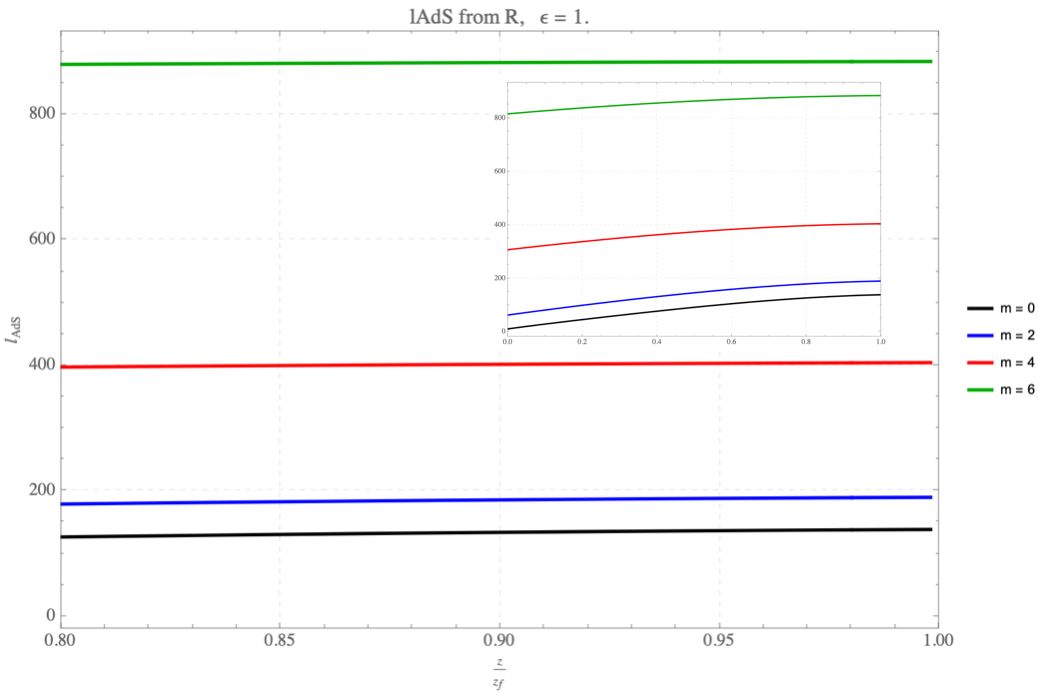}\\
[-0.1cm]\(\text{(a) }l_{AdS} \text{ for } \epsilon=0.1\)&\(\text{(b) }l_{AdS} \text{ for } \epsilon=1\)\end{tabular}\caption{Local AdS radius $l_{\mathrm{AdS}}$ for $\epsilon=0.1$ (a) and $\epsilon=1.0$ (b). The main panels show the region $0.8 \le z/z_f \le 1.0$, and the insets display the corresponding profiles over the full interval $0 \le z/z_f \le 1.0$.}\label{fig:l_ads_profiles}\end{figure}

\begin{figure}[ht]
\centering
\begin{tabular}{cc}
\includegraphics[width=0.47\linewidth]{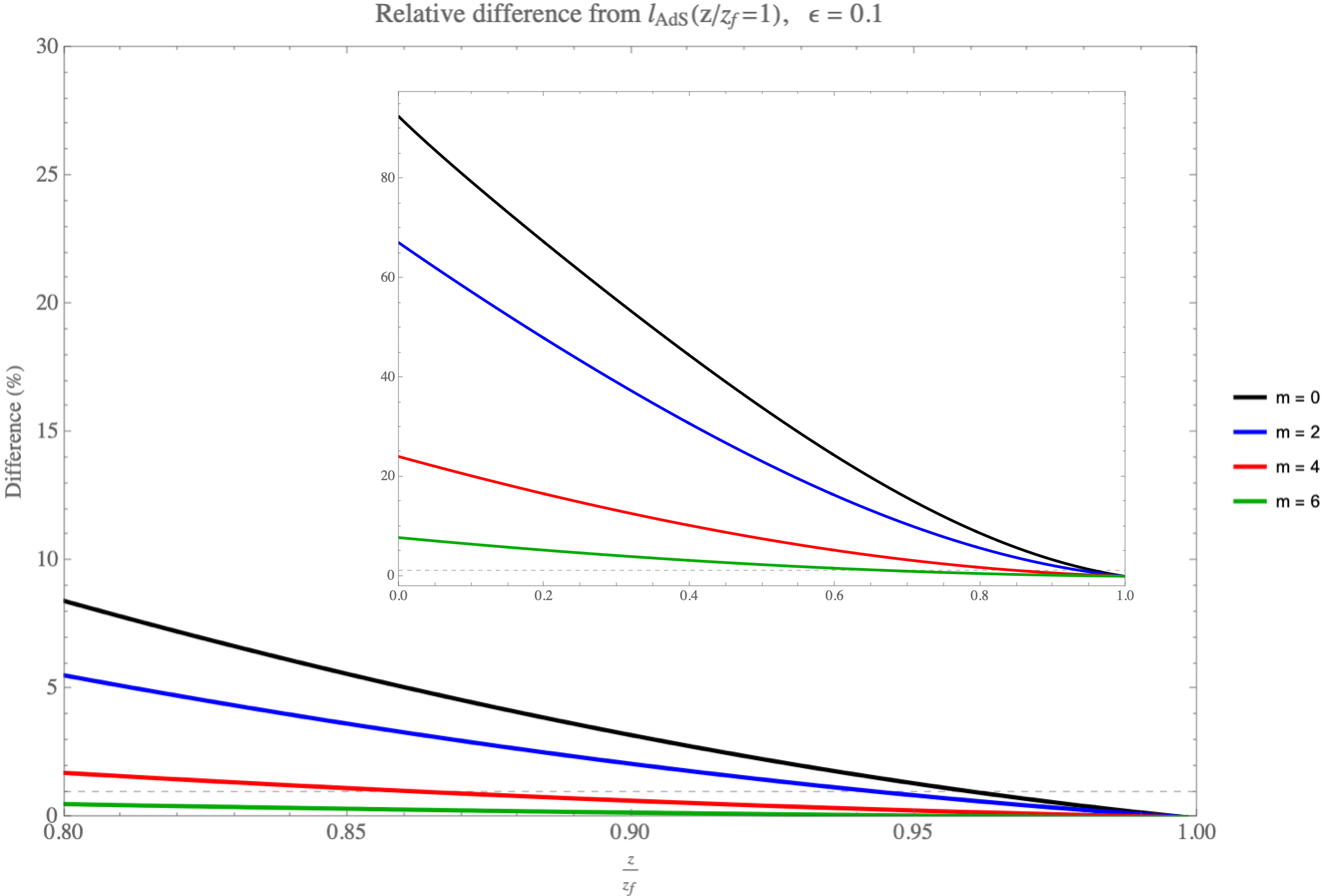}&\includegraphics[width=0.47\linewidth]{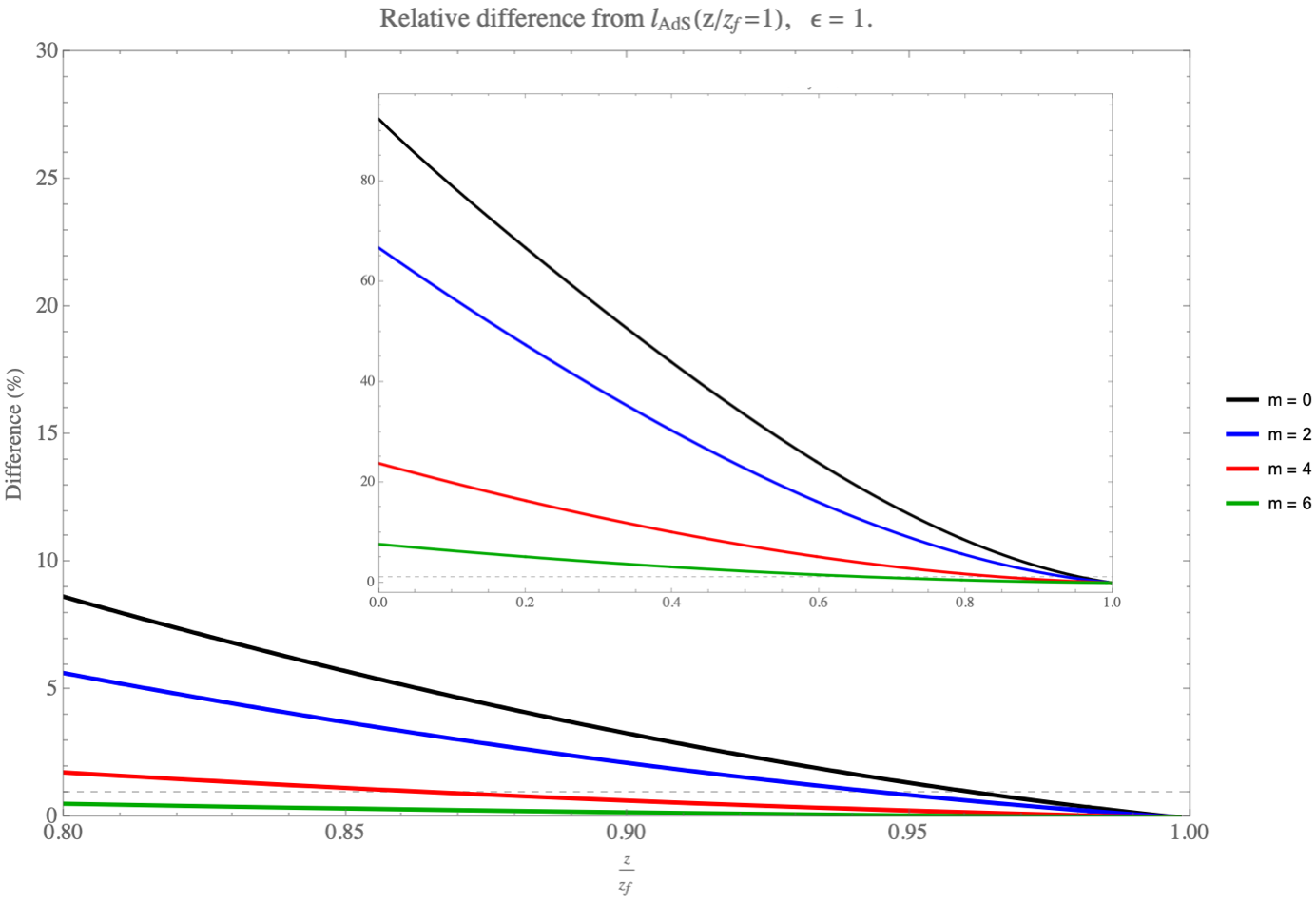}\\
[-0.1cm]\(\text{(a) }\epsilon=0.1\)&\(\text{(b) } \epsilon=1\)\end{tabular}\caption{Relative deviation $\left|l_{\mathrm{AdS}}(z/z_f)-l_{\mathrm{AdS}}(1)\right|/l_{\mathrm{AdS}}(1)$ as a function of $z/z_f$.}\label{fig:cross_over_profiles}\end{figure}

\begin{figure}[ht]
\centering
\begin{tabular}{cc}
\includegraphics[width=0.4\linewidth]{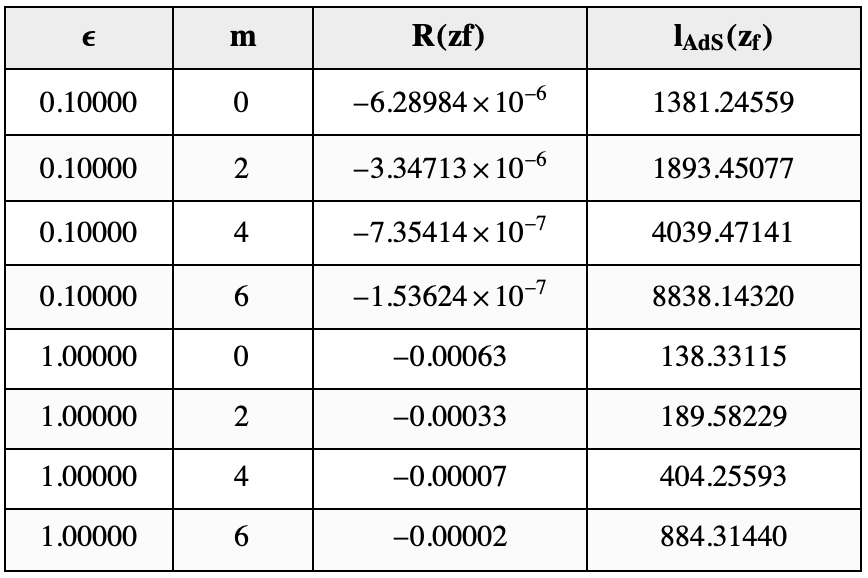}&\includegraphics[width=0.37\linewidth]{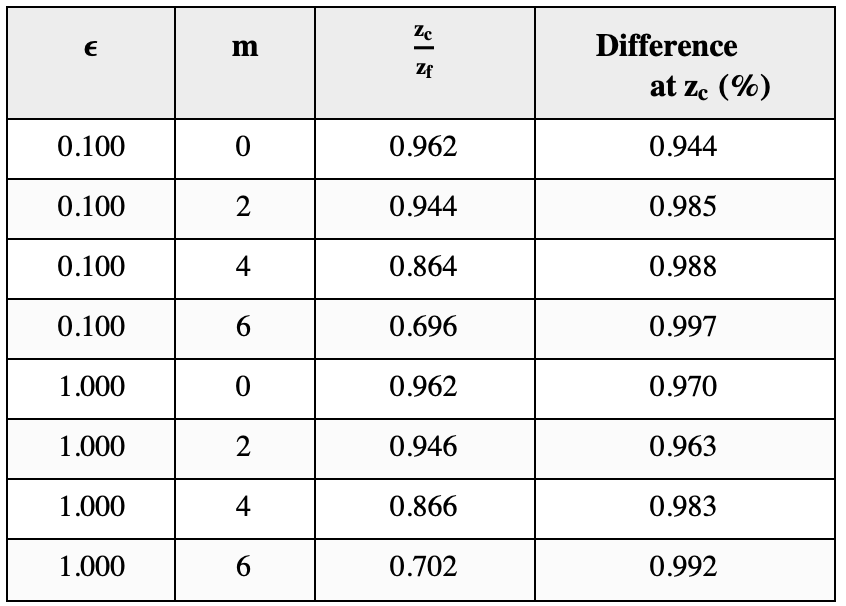}\\
[-0.1cm]\(\text{(a) AdS radius at $z_f$. }\)&\(\text{(b) Crossover position $z_c/z_f$ }\)\end{tabular}\caption{For each pair of $(\epsilon,m)$, (a) reports the infrared value of the AdS radius, while (b) identifies the radial position $z_c$ at which the region with a relative deviation of $l_{\mathrm{AdS}}(z)$ from $l_{\mathrm{AdS}}(z_f)$ below $1\%$ begins.}\label{fig:table}\end{figure}

\begin{figure}[ht]
\centering
\includegraphics[width=0.55\linewidth]{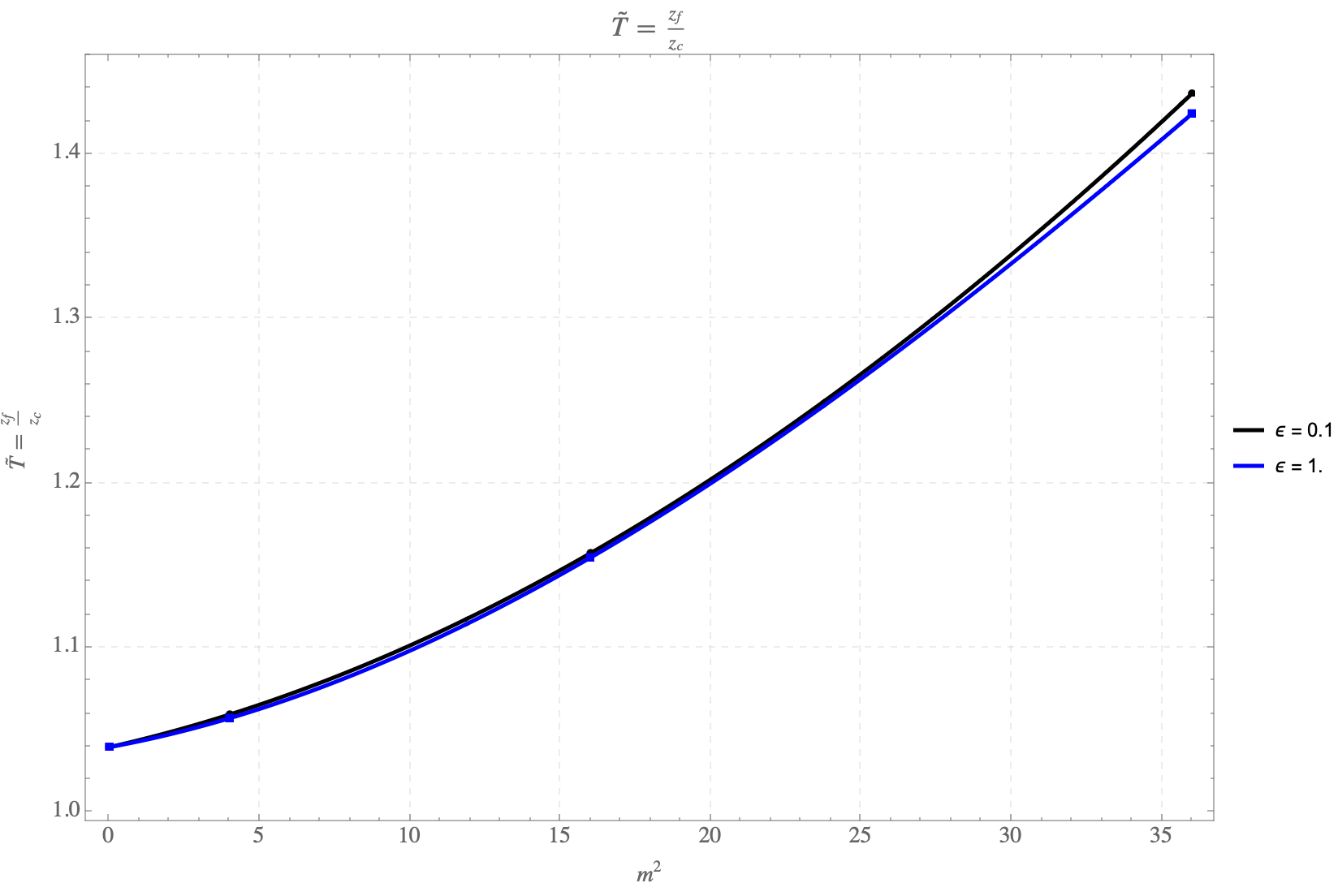}\caption{
Phase diagram in the $(m^2,\widetilde{T})$ plane. Since the temperature scale is inversely proportional to the radial coordinate, $T\propto 1/z$, we define the dimensionless temperature as $\widetilde{T}=z_f/z_c$. The values of $z_f/z_c$ obtained at $m=0,2,4,$ and $6$ are plotted as discrete data points and smoothly interpolated as a function of $m^2$ for different values of $\epsilon$. Here, $z_c$ denotes the onset of the region in which the relative deviation of $l_{\mathrm{AdS}}(z)$ from $l_{\mathrm{AdS}}(z_f)$ is below $1\%$.
}\label{fig:phase_diagram}\end{figure}

In the long-wavelength limit, where the external momentum is set to $p=0$, the last two terms in Eq.~(\ref{Ricci_Scalar_GF}) contain quadratic powers of the internal momenta. In $d=3$, the momentum integration measures introduce additional powers of internal momentum, causing these contributions to exhibit power-law ultraviolet divergences. We therefore define a regularized Ricci scalar by subtracting the last two divergent terms and retaining only the first two terms in Eq.~(\ref{Ricci_Scalar_GF}). The resulting regularized Ricci scalar is shown in Fig.~\ref{fig:ricci_scalar_profiles}.

\noindent With the regularized Ricci scalar, we confirm that the curvature remains negative throughout the bulk. The corresponding AdS radius, defined by $R=-\frac{12}{l_{\mathrm{AdS}}^2}$ in $d=3$, is shown in Fig.~\ref{fig:l_ads_profiles}. For all values of $\epsilon$ and $m$ considered here, $l_{\mathrm{AdS}}(z)$ approaches a constant value toward the IR endpoint, indicating that the emergent geometry becomes asymptotically AdS in the IR. This behavior is quantified in Fig.~\ref{fig:cross_over_profiles} and Fig.~\ref{fig:table}(b), while Fig.~\ref{fig:table}(a) summarizes the endpoint values $l_{\mathrm{AdS}}(z_f)$. The endpoint AdS radius decreases as the mass is reduced, showing that the curvature scale becomes larger and that the geometry evolves from a nearly flat regime toward a more strongly curved AdS regime.

To characterize where this IR AdS behavior sets in, we define the crossover position $z_c$ as the smallest radial coordinate beyond which the relative deviation of $l_{\mathrm{AdS}}(z)$ from its endpoint value $l_{\mathrm{AdS}}(z_f)$ remains below $1\%$. Since the temperature scale is inversely related to the radial coordinate as $T\propto 1/z$, we introduce the dimensionless temperature $\widetilde{T}=z_f/z_c$. Figure~\ref{fig:phase_diagram} shows that $\widetilde{T}$ decreases as $m^2$ approaches its critical value. This behavior is consistent with the generic structure of phase diagrams near a zero-temperature critical point, where the characteristic crossover temperature is suppressed and vanishes at criticality.

\subsubsection{Emergent bulk energy-momentum tensor}
To establish the connection between the emergent geometry and gravitation, the curvature tensors are related to the Einstein field equations:
\bqa && R_{ab}(x,z) - \frac{1}{2} g_{ab}(x,z) R(x,z) = T_{ab}(x,z) , \label{Einstein_Eq} \eqa
where units are chosen such that $8\pi G_N = 1$. 
In this framework, $T_{ab}(x, z)$ represents the effective energy-momentum tensor generated by the collective field fluctuations in the bulk. 
Substituting the curvature expressions (Eqs. (\ref{Ricci_ZZ_GF})--(\ref{Ricci_Scalar_GF})) into the Einstein tensor determines the components of $T_{ab}$.

The radial component, $T_{zz}$, is given by:
\begin{small}\bqa && \begin{split} T_{zz}(x,z)=-\frac{1}{2}\partial_\mu\partial^\mu D(x,z)+\frac{5d-6}{8(d-1)}\frac{\partial_\mu D(x,z)\,\partial^\mu D(x,z)}{D(x,z)}+\frac{d}{8(d-1)}\frac{\left[\partial_zD(x,z)\right]^2}{D^2(x,z)}. \end{split}\nn&& \label{Energy_Momentum_Tensor_ZZ}  \eqa\end{small}
Similarly, off-diagonal and spatial component are expressed as:
\begin{small}
\bqa && T_{z\mu}=\frac{\partial_z\partial_\mu D(x,z)}{2D(x,z)}-\frac{3d-4}{4(d-1)}\frac{\partial_zD(x,z)\,\partial_\mu D(x,z)}{D^2(x,z)}, \label{Energy_Momentum_Tensor_muZ} \eqa
\end{small}
\bqa &&\begin{split} T_{\mu\nu}(x,z)=&-\frac{d-2}{4(d-1)}\frac{\partial_\mu D(x,z)\,\partial_\nu D(x,z)}{D^2(x,z)}+\delta_{\mu\nu}\bigg[\frac{d-2}{8(d-1)}\frac{\partial_\lambda D(x,z)\,\partial^\lambda D(x,z)}{D^2(x,z)}\\
&-\frac{1}{2}\frac{\partial_z^2D(x,z)}{D^2(x,z)}+\frac{7d-8}{8(d-1)}\frac{\left[\partial_zD(x,z)\right]^2}{D^3(x,z)}\bigg].
\end{split}
 \label{Energy_Momentum_Tensor_XX} 
 \eqa

In Lorentzian spacetime, the energy conditions impose local constraints on the stress-energy tensor and provide criteria for assessing whether the matter content exhibits physically reasonable causal and gravitational behavior \cite{Kontou:2020bta}. We adopt the metric signature $(-,+,\cdots,+)$ and consider a $D$-dimensional spacetime. In an orthonormal frame, the stress-energy tensor takes the form
\begin{equation}
T_{\hat{\mu}\hat{\nu}}=\operatorname{diag}\left(\rho,p_1,\ldots,p_{d-1}
\right),\end{equation}
where $\rho$ is the energy density and $p_i$ is the pressure along the $i$th spatial direction.\\
The null energy condition (NEC) requires $T_{\mu\nu}k^\mu k^\nu\geq0$ for every null vector $k^\mu$. In the orthonormal frame, this condition becomes
\begin{equation}
\rho+p_i\geq0
\qquad
\text{for all }i.
\end{equation}
The NEC constrains the stress-energy tensor along null directions and, through the Raychaudhuri equation, is related to the focusing of null geodesic congruences.\\
The weak energy condition (WEC) requires $T_{\mu\nu}u^\mu u^\nu\geq0$ for every timelike vector $u^\mu$. In the orthonormal frame, it is expressed as
\begin{equation}
\rho\geq0, \qquad \rho+p_i\geq0 \qquad \text{for all }i.
\end{equation}
The WEC states that the local energy density measured by any timelike observer is non-negative. Since null vectors arise as limiting cases of timelike vectors, the WEC implies the NEC.\\
The strong energy condition (SEC) requires
\begin{equation}
\left(T_{\mu\nu}-\frac{1}{D-2}Tg_{\mu\nu}\right)u^\mu u^\nu\geq0,
\end{equation}
where $T=g^{\mu\nu}T_{\mu\nu}$ denotes the trace of the stress-energy tensor. In the orthonormal frame, the SEC becomes
\begin{equation}
\rho+p_i\geq0\qquad\text{for all }i,
\end{equation}
together with
\begin{equation}
(D-3)\rho+\sum_{i=1}^{D-1}p_i\geq0.
\end{equation}
Using the Einstein equation, with the cosmological constant absorbed into the effective stress-energy tensor, the SEC may be written geometrically as
\begin{equation}
R_{\mu\nu}u^\mu u^\nu\geq0.
\end{equation}
The SEC is therefore associated with the focusing of timelike geodesic congruences and characterizes the attractive gravitational effect of the effective matter content on neighboring timelike trajectories.\\
The dominant energy condition (DEC) requires the WEC and additionally demands that
\begin{equation}
J^\mu=-T^\mu{}_{\nu}u^\nu
\end{equation}
be a future-directed causal vector for every future-directed timelike vector $u^\mu$. In the orthonormal frame, the DEC is expressed as
\begin{equation}
\rho\geq0,\qquad\rho-\lvert p_i\rvert\geq0\qquad\text{for all }i.
\end{equation}
The DEC requires the energy flux measured by any timelike observer to remain within the local light cone and to be future-directed. It therefore excludes superluminal local energy transport.
These conditions are intrinsically Lorentzian because their definitions rely on timelike and null vectors, as well as on the causal structure determined by the light cone.\\
\begin{figure}[ht]
\centering
\begin{tabular}{cc}
\includegraphics[width=0.47\linewidth]{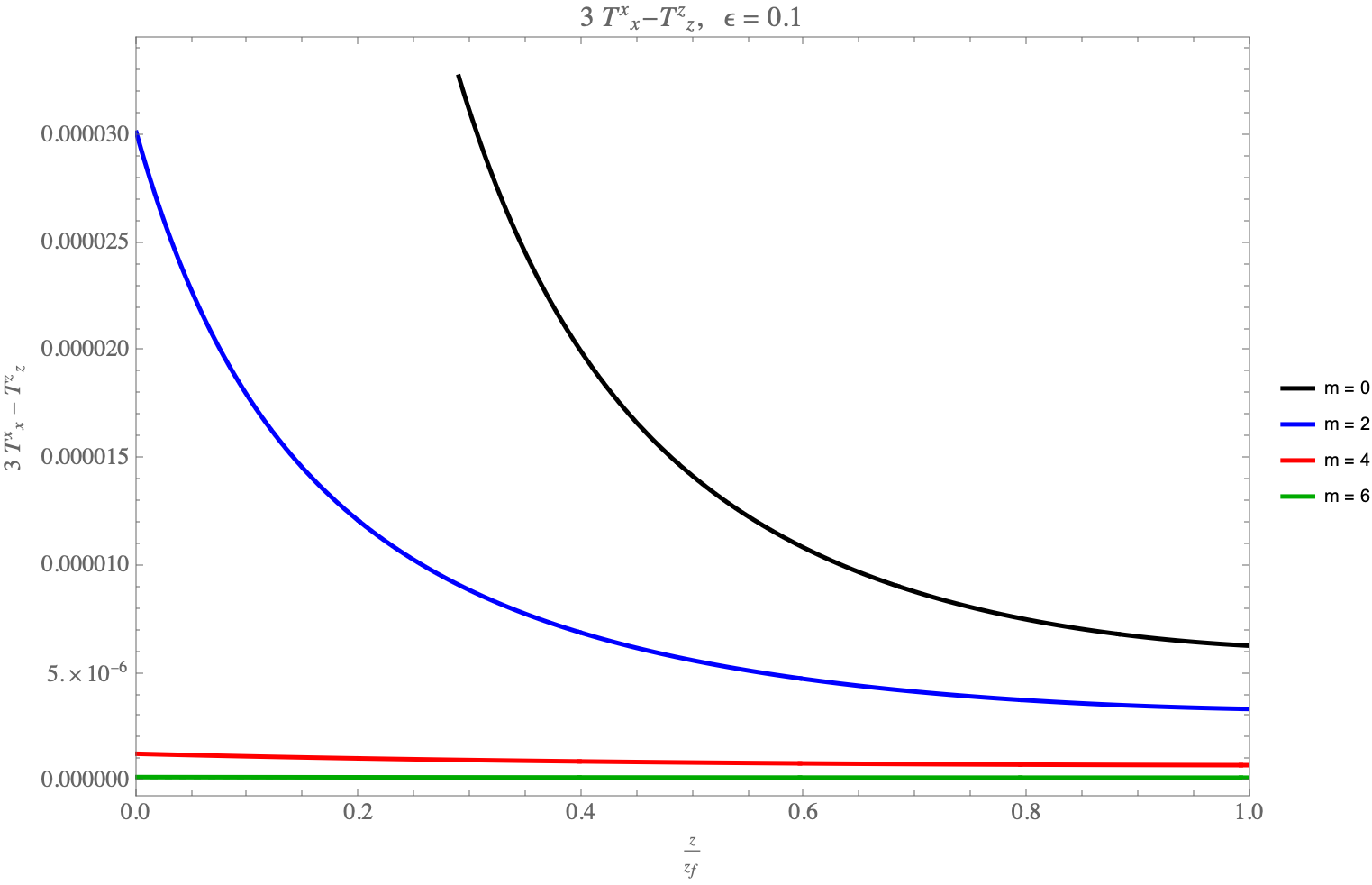}&\includegraphics[width=0.47\linewidth]{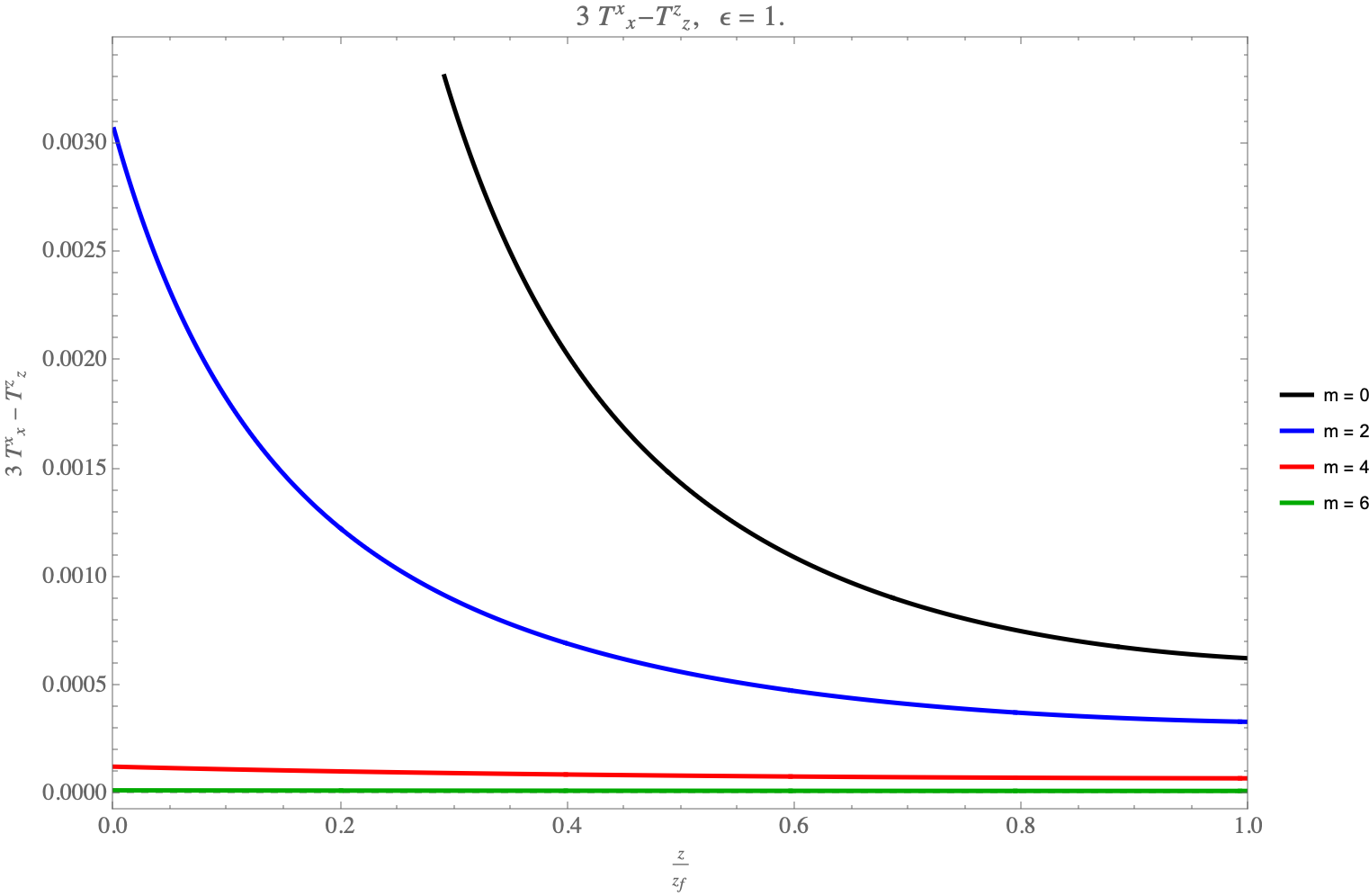}\\
[-0.1cm]\(\text{(a) $\epsilon=0.1$. }\)&\(\text{(b) $\epsilon=1.0$ }\)\end{tabular}\caption{Radial profile of the Euclidean AdS consistency condition, $T^{z}{}_{z}\leq 3T^{x}{}_{x}$, for $d=3$. The inequality is satisfied throughout the radial direction.}\label{fig:Tzz_condition}\end{figure}
\begin{figure}[ht]
\centering
\begin{tabular}{cc}
\includegraphics[width=0.47\linewidth]{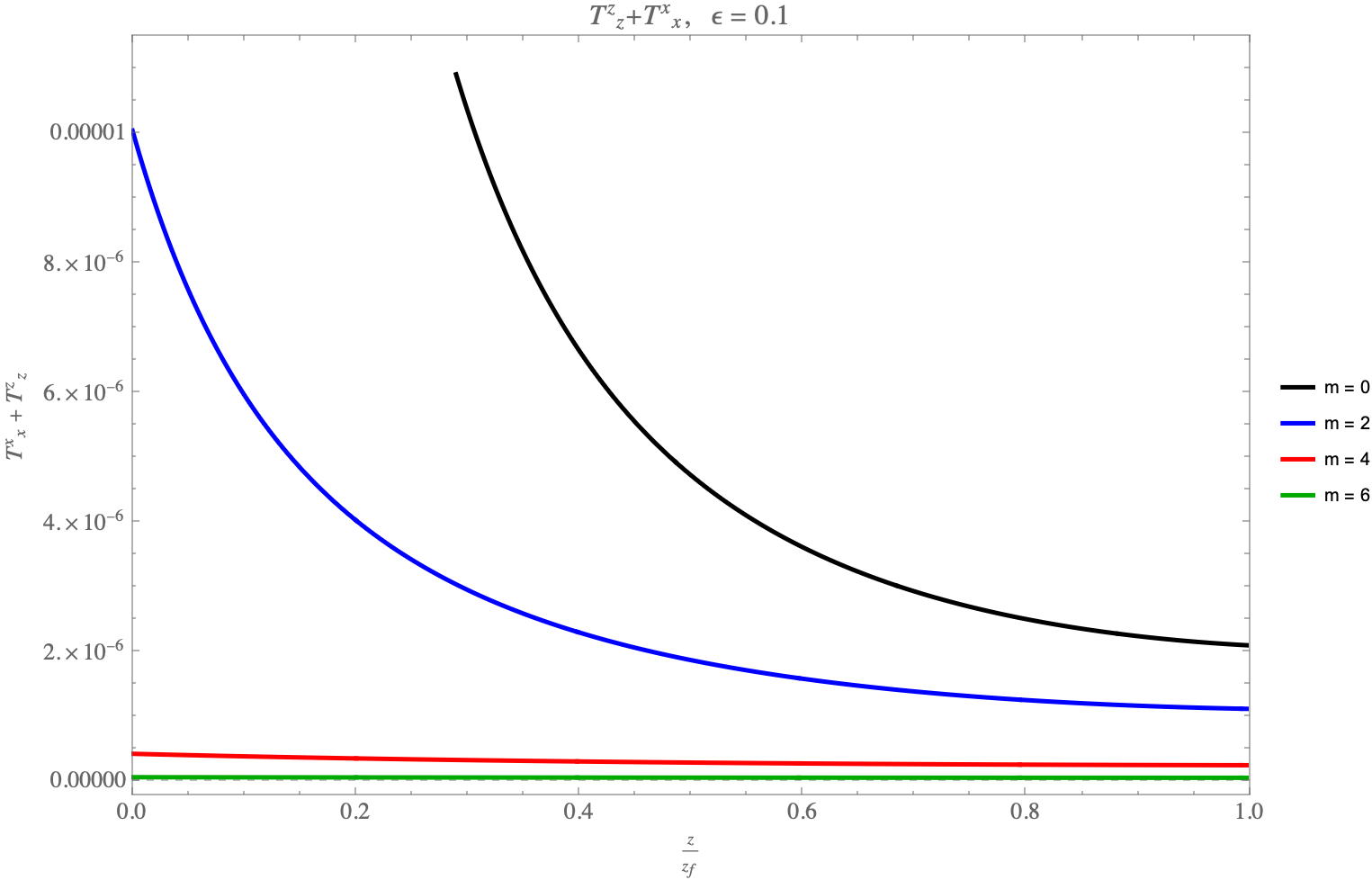}&\includegraphics[width=0.47\linewidth]{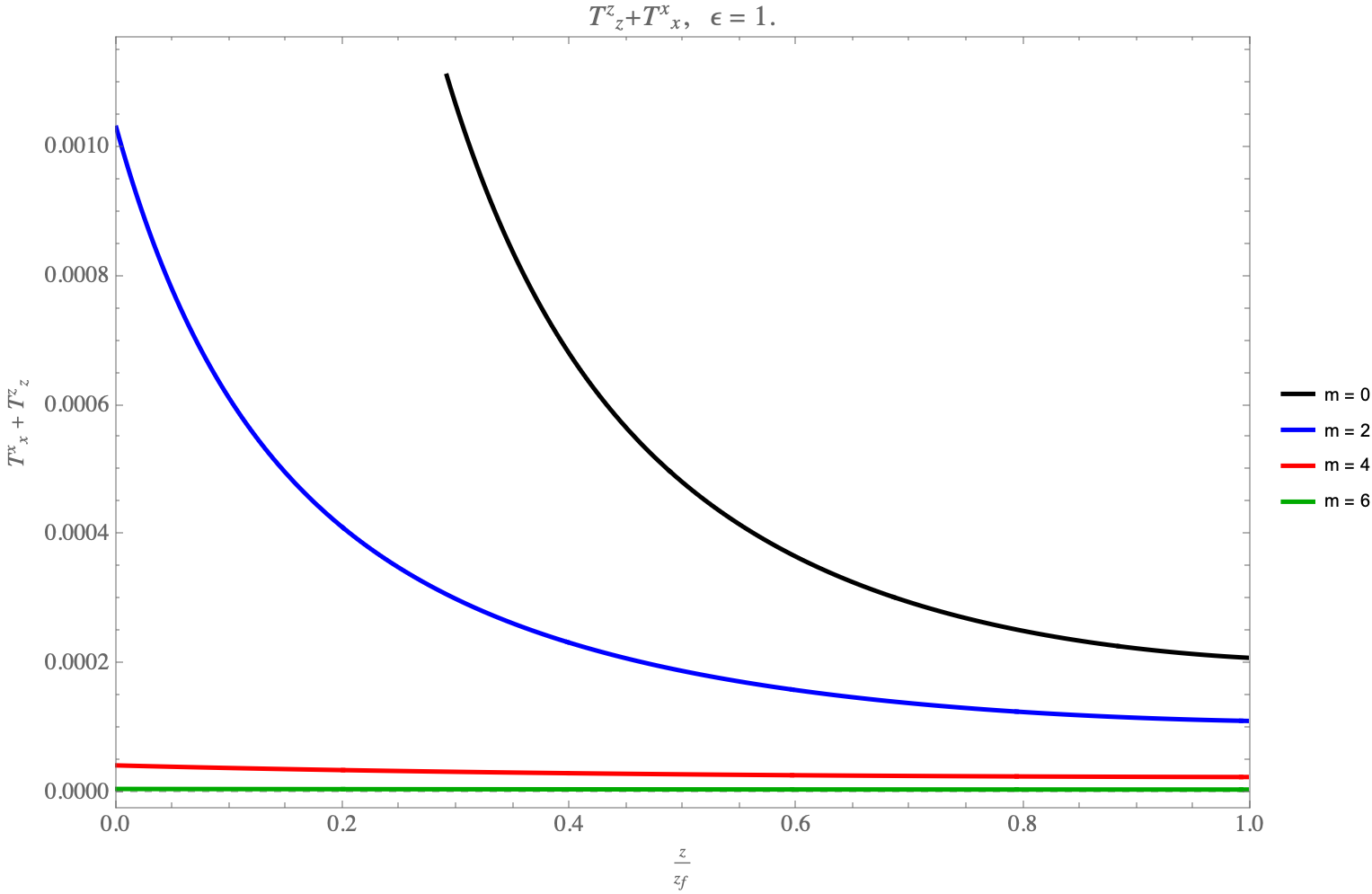}\\
[-0.1cm]\(\text{(a) $\epsilon=0.1$. }\)&\(\text{(b) $\epsilon=1.0$ }\)\end{tabular}\caption{Radial profile of the Euclidean AdS consistency condition, $-T^{x}{}_{x}\leq T^{z}{}_{z}$, for $d=3$. The inequality is satisfied throughout the radial direction.}\label{fig:Txx_condition}\end{figure}

For the Euclidean $\mathrm{AdS}_{d+1}$ geometry, the Ricci tensor and Ricci scalar are given by $R_{ab}=-\frac{d}{l_{\mathrm{AdS}}^{2}}g_{ab}$ and $R=-\frac{d(d+1)}{l_{\mathrm{AdS}}^{2}}$, respectively. Since the Euclidean metric is positive definite, $g_{ab}v^{a}v^{b}>0$ for any nonzero vector $v^{a}$. Using the trace-reversed Einstein equation,
\begin{equation}
T_{ab}-\frac{1}{d-1}Tg_{ab}=R_{ab},
\end{equation}
we obtain
\begin{equation}
\left(T_{ab}-\frac{1}{d-1}Tg_{ab}\right)v^{a}v^{b}
=-\frac{d}{l_{\mathrm{AdS}}^{2}}g_{ab}v^{a}v^{b}
\leq 0.
\end{equation}
The inequality is strict for every nonzero vector. Furthermore, substituting $R_{ab}=-\frac{d}{l_{\mathrm{AdS}}^{2}}g_{ab}$ and $R=-\frac{d(d+1)}{l_{\mathrm{AdS}}^{2}}$ into the Einstein equation gives $T_{ab}=R_{ab}-\frac{1}{2}Rg_{ab}=\frac{d(d-1)}{2l_{\mathrm{AdS}}^{2}}g_{ab}$. Therefore, in an orthonormal frame, $\rho\equiv T_{\hat{0}\hat{0}}=\frac{d(d-1)}{2l_{\mathrm{AdS}}^{2}}>0$. Consequently, the Euclidean AdS geometry in an orthonormal frame satisfies
\begin{equation}
\rho>0,\qquad
T_{aa} - \frac{1}{d-1} T g_{aa} \leq 0.
\end{equation}

In the present calculation, the branch for which the Green function $G(p,z)$ is positive definite is chosen, and the first condition, $\rho>0$, is therefore automatically satisfied. As in the calculation of the regularized Ricci scalar, the spatial-derivative contributions responsible for the ultraviolet divergences are subtracted from the effective energy-momentum tensor. The resulting tensor is diagonal. For $d=3$, the condition applied along the radial direction becomes $T^z_{~z}-\frac{1}{2}\left(3T^x_{~x}+T^z_{~z}\right)\leq 0$, which gives
\begin{equation}
T^z_{~z}\leq 3T^{x}_{~x}.
\end{equation}
Applying the same condition along a spatial direction gives
\begin{equation}
-T^x_{~x}\leq T^{z}_{~z}.
\end{equation}
These two inequalities are satisfied throughout the radial direction, as shown in Figs.~\ref{fig:Tzz_condition} and \ref{fig:Txx_condition}.

\section{Summary and discussion}\label{section4}

In this work, we have constructed a bidirectional holographic dictionary that relates the non-perturbative collective fluctuation sector of the boundary theory to the warping factors of an emergent bulk geometry within the functional renormalization group framework. 
By iterating Wilsonian RG transformations, the RG scale is promoted to the bulk radial coordinate, while the matter propagator $G(x,z)$ and the collective interaction propagator $D(x,z)$ determine the metric components through the holographic dictionary. 
Using the Einstein-space relation, we further extract the effective local AdS radius from the emergent curvature and show that it decreases continuously as the mass gap is reduced. 
These results demonstrate that the emergent geometry evolves continuously from an approximately flat spacetime toward the $\mathrm{AdS}_{d+1}$ vacuum as the boundary theory approaches criticality.

The present analysis is performed in the regime where the interaction propagator is approximated by its coupling-dominated contribution, reducing the bulk dynamics to a tractable second-order differential equation. 
Although this approximation captures the continuous evolution of the bulk metric and curvature, it does not incorporate the complete non-linear backreaction generated by the full collective interaction
$D(x,z)=\left[\frac{1}{u}+G^2(x,z)\right]^{-1}\delta(x)$, whose polarization function is determined by the non-perturbative $GG$ channel. A complete treatment of the non-perturbative $GG$ channel may also clarify the relation between collective four-point fluctuations and the emergent bulk geometry, analogous to the role of four-point conformal blocks in encoding AdS background geometries in conformal-bootstrap approaches \cite{Fitzpatrick:2014vua}.
Solving the full coupled system therefore remains an important direction for obtaining the complete bulk dynamics beyond the present approximation. Another important consistency test is to compare the effective energy-momentum tensor derived here with that obtained from the Einstein--Hilbert bulk action in Ref.~\cite{Nonperturbative_Wilson_RG}. Agreement between the two would provide independent evidence that the AdS geometry extracted from the Ricci scalar and the effective source determined by the Einstein equations describe the same emergent gravitational background.

An immediate extension of the present framework is to finite-temperature field theories. 
In this construction, the inverse temperature $\beta$ is related to the final RG scale $z_f$. 
A self-consistent solution of the full coupled radial equations may generate thermal bulk geometries with non-trivial horizon structures. 
For such a solution to constitute a consistent holographic black-hole geometry, it must satisfy the Euclidean thermal identification $\tau\equiv\tau+\beta$ together with the matching condition $\beta=\beta_H$, where $\beta_H=1/T_H$ is the inverse Hawking temperature. 
The same construction must also reproduce the Bekenstein--Hawking area law for the horizon entropy and ensure the cancellation of logarithmic ultraviolet divergences in the boundary effective action, leaving finite renormalized thermodynamic quantities. 
Establishing these thermal, geometric, and renormalization conditions simultaneously would provide a non-perturbative derivation of black-hole thermodynamics directly from the underlying RG flow.
We leave these developments for future work.

\begin{acknowledgments}
K.-S. K. was supported by the Ministry of Education, Science, and Technology (Grant No. RS-2024-00337134) of the National Research Foundation of Korea (NRF). K.-S. K. appreciate insightful discussions with D. Mukherjee, A. Mitra, Y. Choun, J.-M. Bok, S.-J. Yoo, L. J. F. Sese, S. Park, and P. Salgado-Rebolledo.
\end{acknowledgments}

\appendix

\section{Derivation of the holographic dual effective field theory in the functional RG analysis}\label{appendixA}

To derive the holographic dual effective field theory, an auxiliary field $\psi_\alpha(x)$ is introduced into the $O(N)$ vector model. 
The partition function is expressed as:
\bqa && \begin{split}Z = \int D \psi_{\alpha}(x) D \phi_{\alpha}(x) D \varphi(x) \exp\Big[ - \int d^{d} x &\Big\{ \phi_{\alpha}(x) \Big( - \partial_{\mu}^{2} + m^{2} - i \varphi(x) \Big) \phi_{\alpha}(x) \\
&+ M^{2} \psi_{\alpha}^{2}(x) + \frac{N}{2 u} \varphi^{2}(x) \Big\} \Big] . \end{split}\eqa
Here, $\psi_\alpha(x)$ has a mass $M$ and is used to define the scale for the RG transformation. 
A field transformation mixing $\phi_\alpha(x)$ and $\psi_\alpha(x)$ separates the high-energy and low-energy modes. 
This approach determines the Jacobian factors from the field transformation, which enter the kinetic and potential terms in the emergent $(d+1)$-dimensional bulk action.

To implement the Wilsonian RG transformation, the fields are decomposed into low-energy and high-energy fluctuations via the following linear transformation:
\bqa && \phi_{\alpha}(x) \longrightarrow \phi_{\alpha}(x) + \Phi_{\alpha}(x) , ~~~~~ \psi_{\alpha}(x) \longrightarrow c_{\phi} \phi_{\alpha}(x) + c_{\Phi} \Phi_{\alpha}(x) , \eqa
where $\phi_\alpha(x)$ and $\Phi_\alpha(x)$ denote the low-energy and high-energy degrees of freedom, respectively. 
The transformation coefficients, $c_{\phi}$ and $c_{\Phi}$, are given by:
\bqa && c_{\phi} = \frac{- \partial_{\mu}^{2} + m^{2} - i \varphi(x)}{\mu M} , ~~~~~ c_{\Phi} = - \frac{\mu}{M} , ~~~~~ \mu = \frac{\sqrt{- \partial_{\mu}^{2} + m^{2} - i \varphi(x)}}{\sqrt{e^{2 \alpha d z} - 1}} . \eqa
These coefficients are set by the condition that the cross-terms between the low-energy and high-energy modes vanish in the transformed action. 
Consequently, the quadratic part of the action is diagonalized as:
\begin{small}\bqa && \phi_{\alpha}(x) \Big( - \partial_{\mu}^{2} + m^{2} - i \varphi(x) \Big) \phi_{\alpha}(x) + M^{2} \psi_{\alpha}^{2}(x) \nn && = e^{2 \alpha d z} \phi_{\alpha}(x) \Big( - \partial_{\mu}^{2} + m^{2} - i \varphi(x) \Big) \phi_{\alpha}(x) + \frac{e^{2 \alpha d z}}{e^{2 \alpha d z} - 1} \Phi_{\alpha}(x) \Big( - \partial_{\mu}^{2} + m^{2} - i \varphi(x) \Big) \Phi_{\alpha}(x) .\hspace{0.8em}
 \eqa\end{small}
This decoupling enables the high-energy modes $\Phi_\alpha(x)$ to be integrated out, yielding the effective action for the low-energy modes $\phi_\alpha(x)$.

Substituting the diagonalized action into the partition function yields:
\bqa && Z = \int D \Phi_{\alpha}(x) D \phi_{\alpha}(x) D \varphi(x) ~ \mathcal{J}[i \varphi(x)] \nn && \exp\Big[ - \int d^{d} x \Big\{ \phi_{\alpha}(x) \Big( - \partial_{\mu}^{2} + m^{2} - i \varphi(x) \Big) \phi_{\alpha}(x) \nn && + \frac{1}{e^{2 \alpha d z} - 1} \Phi_{\alpha}(x) \Big( - \partial_{\mu}^{2} + m^{2} - i \varphi(x) \Big) \Phi_{\alpha}(x) + \frac{N}{2 u} \varphi^{2}(x) \Big\} \Big] . \eqa
The fields are rescaled as follows:
\bqa && \phi_{\alpha}(x) \longrightarrow e^{- \alpha d z} \phi_{\alpha}(x) , ~~~~~ \Phi_{\alpha}(x) \longrightarrow e^{- \alpha d z} \Phi_{\alpha}(x) , \eqa 
where $\alpha$ is a constant parameter ($\partial_\mu \alpha = 0$). 
The change of variables from $(\phi, \psi)$ to $(\phi, \Phi)$ defined in Eq. (\ref{Jacobian_Matrix}):
\bqa && \begin{pmatrix} \phi(x) \\ \psi(x) \end{pmatrix} = \begin{pmatrix} 1 & 1 \\ c_{\phi} & c_{\Phi} \end{pmatrix} \begin{pmatrix} \phi(x) \\ \Phi(x) \end{pmatrix} \label{Jacobian_Matrix} \eqa
yields the Jacobian factor $\mathcal{J}[i\varphi(x)]$, which represents the transformation of the functional measure:
\bqa && \mathcal{J}[i \Sigma(x)] = \exp\Big\{ \int d^{d} x 2 \alpha d z \frac{N}{4} \ln \Big( - \partial_{\mu}^{2} + m^{2} - i \varphi(x) \Big) \Big\} . \label{Jacobian_A} \eqa

After performing the Gaussian integration over the high-energy matter fields $\Phi_\alpha(x)$ and incorporating the Jacobian factor, we obtain the effective action for the low-energy modes $\phi_\alpha(x)$:
\bqa && Z = \int D \phi_{\alpha}(x) D \varphi(x) \exp\Big[ - \int d^{d} x \Big\{ \phi_{\alpha}(x) \Big( - \partial_{\mu}^{2} + m^{2} - i \varphi(x) \Big) \phi_{\alpha}(x) + \frac{N}{2 u} \varphi^{2}(x) \nn && + 2 \alpha d z \frac{N}{4} \ln \Big( - \partial_{\mu}^{2} + m^{2} - i \varphi(x) \Big) \Big\} \Big] . \label{After_RG_Matter} \eqa
In Eq. (\ref{After_RG_Matter}), the high-energy matter fields are integrated out, but the field $\varphi(x)$ remains defined at a single scale. 
To define the holographic dimension, the fluctuations of the collective field during this RG step are included.

Next, the RG transformation is applied to the collective field $\varphi(x)$. 
The field $\varphi(x)$ in Eq. (\ref{After_RG_Matter}) is replaced with $\varphi^{(0)}(x)$, and an auxiliary collective field $\tilde{\varphi}^{(0)}(x)$ is introduced:
\bqa && Z = \int D \phi_{\alpha}(x) D \tilde{\varphi}^{(0)}(x) D \varphi^{(0)}(x) \exp\Big[ - \int d^{d} x \Big\{ \phi_{\alpha}(x) \Big( - \partial_{\mu}^{2} + m^{2} - i \varphi^{(0)}(x) \Big) \phi_{\alpha}(x) \nn && + \frac{N}{2 u} \varphi^{(0) 2}(x) + \frac{N}{2 \tilde{u}} \tilde{\varphi}^{(0) 2}(x) + 2 \alpha d z \frac{N}{4} \ln \Big( - \partial_{\mu}^{2} + m^{2} - i \varphi^{(0)}(x) \Big) \Big\} \Big] . \eqa
The addition of $\tilde{\varphi}^{(0)}(x)$ with the coupling $\tilde{u}$ enables the decomposition of the collective field into low-energy and high-energy degrees of freedom.

The low-energy and high-energy fluctuations within the collective sector are separated via the following linear transformation:
\bqa && \varphi^{(0)}(x) \longrightarrow \varphi^{(0)}(x) + \tilde{\varphi}^{(1)}(x) , ~~~~~ \tilde{\varphi}^{(0)}(x) \longrightarrow c_{\varphi} \varphi^{(0)}(x) + c_{\tilde{\varphi}} \tilde{\varphi}^{(1)}(x) , \eqa
where $\varphi^{(0)}(x)$ and $\tilde{\varphi}^{(1)}(x)$ denote the low-energy and high-energy degrees of freedom of the collective field, respectively. 
The transformation coefficients, $c_{\varphi}$ and $c_{\tilde{\varphi}}$, are given by:
\bqa && c_{\varphi} = \frac{\sqrt{\tilde{u}}}{\mu u} , ~~~~~ c_{\tilde{\varphi}} = - \mu \sqrt{\tilde{u}} , ~~~~~ \mu = \frac{1}{\sqrt{u} \sqrt{e^{2 \beta d z} - 1}} . \eqa
These coefficients are set by the condition that the cross-terms between the low-energy and high-energy modes vanish in the transformed action. 
Consequently, the quadratic part of the collective field action is diagonalized as:
\bqa && \frac{N}{2 u} \varphi^{(0) 2}(x) + \frac{N}{2 \tilde{u}} \tilde{\varphi}^{(0) 2}(x) = e^{2 \beta d z} \frac{N}{2 u} \varphi^{(0) 2}(x) + \frac{e^{2 \beta d z}}{e^{2 \beta d z} - 1} \frac{N}{2 u} \tilde{\varphi}^{(1) 2}(x) , \eqa
where $\beta$ serves as a control parameter for the RG transformation of the collective field. 
This diagonalization allows the high-energy collective fluctuations $\tilde{\varphi}^{(1)}(x)$ to be integrated out.

Substituting the diagonalized expression into the partition function separates the modes of the collective field. 
Since the Jacobian for this field transformation is equal to unity, it is omitted. 
The resulting expression is given by:
\bqa && Z = \int D \phi_{\alpha}(x) D \tilde{\varphi}^{(1)}(x) D \varphi^{(0)}(x) \nn && \exp\Big[ - \int d^{d} x \Big\{ \phi_{\alpha}(x) \Big( - \partial_{\mu}^{2} + m^{2} - i \varphi^{(0)}(x) - i \tilde{\varphi}^{(1)}(x) \Big) \phi_{\alpha}(x) + \frac{N}{2 u} \varphi^{(0) 2}(x)  \nn &&+ \frac{1}{2 \beta d z} \frac{N}{2 u} \tilde{\varphi}^{(1) 2}(x) + 2 \alpha d z \frac{N}{4} \ln \Big( - \partial_{\mu}^{2} + m^{2} - i \varphi^{(0)}(x) - i \tilde{\varphi}^{(1)}(x) \Big) \Big\} \Big] , \label{RG_Varphi} \eqa
The collective fields are rescaled as follows:
\bqa && \varphi^{(0)}(x) \longrightarrow e^{- \beta d z} \varphi^{(0)}(x) , ~~~~~ \tilde{\varphi}^{(1)}(x) \longrightarrow e^{- \beta d z} \tilde{\varphi}^{(1)}(x) , \eqa
where the approximation $e^{2\beta dz} - 1 \approx 2\beta dz$ has been applied for the scale parameter $dz$. 
This rescaling completes the transformation for the collective sector along the radial coordinate.

To perform the functional integration over the high-energy collective modes $\tilde{\varphi}^{(1)}(x)$, the logarithmic term in Eq. (\ref{RG_Varphi}) is expanded up to second order in $\tilde{\varphi}^{(1)}(x)$:
\bqa && 2 \alpha d z \frac{N}{4} \ln \Big( - \partial_{\mu}^{2} + m^{2} - i \varphi^{(0)}(x) - i \tilde{\varphi}^{(1)}(x) \Big) \nn && \approx 2 \alpha d z \frac{N}{4} \ln \Big( - \partial_{\mu}^{2} + m^{2} - i \varphi^{(0)}(x) \Big) - 2 \alpha d z \frac{N}{4} \Big( - \partial_{\mu}^{2} + m^{2} - i \varphi^{(0)}(x) \Big)^{-1} i \tilde{\varphi}^{(1)}(x) \nn && - 2 \alpha d z \frac{N}{8} i \tilde{\varphi}^{(1)}(x) \Big( - \partial_{\mu}^{2} + m^{2} - i \varphi^{(0)}(x) \Big)^{-1} \Big( - \partial_{\mu}^{2} + m^{2} - i \varphi^{(0)}(x) \Big)^{-1} i \tilde{\varphi}^{(1)}(x) . \label{Log_Expansion} \eqa
The linear term in $\tilde{\varphi}^{(1)}(x)$ shifts the field configuration, while the quadratic term modifies the propagator of the collective sector, which enters the RPA-corrected kinetic structure in the bulk theory.

Substituting Eq. (\ref{Log_Expansion}) into the action, the Gaussian integration over the field $\tilde{\varphi}^{(1)}(x)$ is performed. 
To decouple the interactions in the fluctuation sector, a Hubbard-Stratonovich transformation is applied by introducing a conjugate field $\varphi^{(1)}(x)$. 
This yields the final expression for the partition function:
\begin{small} \bqa &&Z = \int D \phi_{\alpha}(x) D \varphi^{(1)}(x) D \varphi^{(0)}(x) \exp\Big[ - \int d^{d} x \Big\{ \phi_{\alpha}(x) \Big( - \partial_{\mu}^{2} + m^{2} - i \varphi^{(1)}(x) \Big) \phi_{\alpha}(x) + \frac{N}{2 u} \varphi^{(0) 2}(x) \nn && + \frac{1}{2 \beta d z} \frac{N}{2} \Big( \varphi^{(1)}(x) - \varphi^{(0)}(x) \Big) \Big\{ \frac{1}{u} + \alpha \beta (d z)^{2} \Big( - \partial_{\mu}^{2} + m^{2} - i \varphi^{(0)}(x) \Big)^{-2} \Big\} \Big( \varphi^{(1)}(x) - \varphi^{(0)}(x) \Big) \nn && - 2 \alpha d z \frac{N}{4} \Big( - \partial_{\mu}^{2} + m^{2} - i \varphi^{(0)}(x) \Big)^{-1} i \Big( \varphi^{(1)}(x) - \varphi^{(0)}(x) \Big) + 2 \alpha d z \frac{N}{4} \ln \Big( - \partial_{\mu}^{2} + m^{2} - i \varphi^{(0)}(x) \Big) \Big\} \Big] , \nn&& \eqa \end{small}
where a field shift $\varphi^{(1)}(x) \rightarrow \varphi^{(1)}(x) - \varphi^{(0)}(x)$ is absorbed. 
The coupling term $\phi_\alpha(x) (-\partial_\mu^2 + m^2 - i\varphi^{(1)}(x)) \phi_\alpha(x)$ shows that the collective field $\varphi^{(0)}(x)$ at scale $z$ evolves into $\varphi^{(1)}(x)$ at scale $z+dz$. 
This relation relates the discrete transformation steps to the continuous holographic bulk action presented in the main text.

\section{Boundary conditions}\label{appendixB}

To derive the boundary conditions for the field $\varphi(x, z)$, the partition function is expressed in terms of $\varphi(x, z)$ and its conjugate momentum $\Pi_\varphi(x, z)$. 
Collecting the contributions from the RG process, including the Jacobian factors and the boundary terms at $z=0$ and $z=z_f$, yields the following partition function:
\begin{small}\bqa && Z = \int D \Pi_{\varphi}(x,z) D \varphi(x,z) \exp\Bigg[ - N \int d^{d} x \Bigg\{ \frac{1 + \varepsilon\alpha}{2} \ln G^{-1}(x,z_{f}) - \frac{\varepsilon\alpha}{2} \ln G^{-1}(x,0) + \frac{1}{2 u} \varphi^{2}(x,0) \Bigg\} \nn && - N \int_{0}^{z_{f}} d z \int d^{d} x \Bigg\{ \beta \Pi_{\varphi}(x,z) D(x,z) \Pi_{\varphi}(x,z) -i \Pi_{\varphi}(x,z) \Big( \partial_{z} \varphi(x,z) \Big) + \frac{\alpha}{2} \ln G^{-1}(x,z) \Bigg\} \Bigg] . \label{Partition_Function_Hamiltonian_Form} \eqa \end{small}
In this formulation, the first integral in the exponent represents the boundary action at the UV ($z=0$) and IR ($z=z_f$) limits. 
The second integral represents the bulk action in a first-order (Hamiltonian) framework, where $\Pi_\varphi(x, z)$ acts as the momentum field along the radial $z$-direction. 
The terms containing $\ln G^{-1}$ represent the quantum corrections integrated out during the RG flow. 
This structure defines the functional relation for the conjugate momenta and the boundary constraints.

The relation between the evolution of the field and its conjugate momentum is determined by evaluating the stationary point of the bulk action in Eq. (\ref{Partition_Function_Hamiltonian_Form}). 
Taking the functional variation with respect to the momentum field $\Pi_\varphi(x, z)$ yields the constitutive relation:
\bqa && \Pi_{\varphi}(x,z) = \frac{i}{2 \beta} D^{-1}(x,z) \partial_{z} \varphi(x,z) . \label{Canonical_Momentum} \eqa
Equation (\ref{Canonical_Momentum}) defines $\Pi_\varphi(x, z)$ as the conjugate momentum associated with the field $\varphi(x, z)$ along the radial dimension, representing the flow of collective interactions modulated by the propagator $D(x, z)$. 
Substituting this relation into Eq. (\ref{Partition_Function_Hamiltonian_Form}) yields the second-order kinetic term $(\partial_{z} \varphi)^{2}$ of the bulk Lagrangian in Eq. (\ref{Bulk_Lagrangian}).

By combining the initial boundary terms in Eq. (\ref{Partition_Function_Hamiltonian_Form}) with the surface terms generated during the variation of the bulk action, the total effective action $S_{eff}$ is defined as:
\begin{small}
\bqa && \begin{split}S_{eff} = N \int d^{d} x \Bigg\{ &\frac{1 + \varepsilon\alpha}{2} \ln G^{-1}(x,z_{f}) - \frac{\varepsilon\alpha}{2} \ln G^{-1}(x,0) + \frac{1}{2 u} \varphi^{2}(x,0) -i\Pi_{\varphi}(x,z_{f}) \varphi(x,z_{f}) \\
&+i\Pi_{\varphi}(x,0) \varphi(x,0) \Bigg\} . \label{Boundary_Action} \end{split}\eqa
\end{small}
In this expression, the terms $\Pi_\varphi(x, z_f)\varphi(x, z_f)$ and $-\Pi_\varphi(x, 0)\varphi(x, 0)$ are generated by the integration by parts of the bulk action. 
This effective action serves as the generating functional for the correlation functions of the theory, where the relation between the momentum $\Pi_{\varphi}$ and the field $\varphi$ at the boundaries determines the state of the system. 
Requiring this action to be stationary with respect to field variations at $z=0$ and $z=z_f$ yields the boundary conditions for the bulk equations of motion.

The boundary conditions are determined by requiring the variation of the total effective action $S_{eff}$ in Eq. (\ref{Boundary_Action}) with respect to the fields $\varphi(x, z_f)$ and $\varphi(x, 0)$ to vanish. 
At the IR boundary ($z = z_f$), this stationarity condition relates the momentum to the matter-field propagator:
\bqa && \Pi_{\varphi}(x,z_{f}) = -\frac{1 + \varepsilon\alpha}{2} G(x,z_{f}) = \frac{i}{2 \beta} D^{-1}(x,z_{f}) \partial_{z_{f}} \varphi(x,z_{f}) \nn&&\longrightarrow \partial_{z_{f}} \varphi(x,z_{f}) = i \beta (1 + \varepsilon\alpha) D(x,z_{f}) G(x,z_{f}) . \label{IR_BC_C} \eqa
This relation specifies the derivative of the field at the infrared boundary in terms of the fluctuations of the matter field. 
Similarly, at the UV boundary ($z = 0$), the variation with respect to $\varphi(x, 0)$ yields:
\bqa && \Pi_{\varphi}(x,0) = - \frac{\varepsilon\alpha}{2} G(x,0) + \frac{i}{u} \varphi(x,0) = \frac{i}{2 \beta} D^{-1}(x,0) [\partial_{z} \varphi(x,z)]_{z = 0} \nn && \longrightarrow [\partial_{z} \varphi(x,z)]_{z = 0} = \frac{2 \beta}{u} D(x,0) \varphi(x,0) + i \varepsilon\alpha \beta D(x,0) G(x,0) . \label{UV_BC_C} \eqa
Equation (\ref{UV_BC_C}) relates the microscopic interaction $u$ and the boundary field $\varphi(x, 0)$ to the initial data for the radial flow. 
Substituting the form of the canonical momentum into these variational results yields the boundary conditions presented in Eqs. (\ref{IR_BC}) and (\ref{UV_BC}) of the main text. 
This derivation defines the dual theory as a boundary value problem in $d+1$ dimensions.

\section{Functional RG equation}\label{appendixC}

To describe the scale-dependent evolution of the system, the functional renormalization group (RG) equation is derived. 
The infinitesimal transformation of the partition function is generated by the RG Hamiltonian operator $\mathcal{H}_{RG}$. 
Accounting for variations in both the field $\phi(x, z)$ and the field $\varphi(x, z)$ during an infinitesimal RG step, this operator is defined as:
\bqa && \begin{split}\mathcal{H}_{RG} = N \int d^{d} x \Bigg\{ &- \beta \frac{\delta}{\delta \varphi(x,z)} D(x,z) \frac{\delta}{\delta \varphi(x,z)} + \frac{\delta}{\delta \varphi(x,z)} \varphi(x,z) \\
&- \alpha \frac{\delta}{\delta \phi(x,z)} G(x,z) \frac{\delta}{\delta \phi(x,z)} + \frac{\delta}{\delta \phi(x,z)} \phi(x,z) \Bigg\} .\end{split}  \label{RG_Hamiltonian} \eqa
In this expression, the first two terms represent the RG flow of the collective sector, whereas the latter two terms represent the evolution of the matter sector. 
The terms quadratic in the functional derivatives, which are proportional to the propagators $D$ and $G$, account for the integration of high-energy fluctuations in the field space.

The terms linear in the functional derivatives represent the drift or rescaling of the fields as the RG flow evolves toward the infrared. 
This functional operator $\mathcal{H}_{RG}$ acts on the field probability distribution $P[\varphi(x, z), \phi(x, z)]$, describing the evolution of the effective action along the radial dimension $z$. 
This framework defines the relations for the bulk dynamics discussed in the main text.

The RG Hamiltonian defined in Eq. (\ref{RG_Hamiltonian}) is determined by the background field configurations through the following Green's functions. 
The matter-field propagator $G(x, z)$, which describes the fluctuations of the fields $\phi(x, z)$ at a given scale $z$, is defined by the relation:
\bqa && \Big( - \partial_{\mu}^{2} + m^{2} - i \varphi(x,z) \Big) G(x,z) = \delta^{(d)}(x) . \eqa 
This Green's function represents the response of the matter fields to the local configuration of the field $\varphi(x, z)$. 
Similarly, the propagator $D(x, z)$ for the collective field, which includes the Random Phase Approximation (RPA) corrections from the matter-field fluctuations, is given by:
\bqa && \Big( \frac{1}{u} + \alpha \beta \varepsilon^{2} G^{2}(x,z) \Big) D(x,z) = \delta^{(d)}(x) . \eqa
This relation shows that the interaction between collective modes is modified by the squared propagator of the matter fields, $G^2(x, z)$. 
Within the functional RG framework, these two equations specify the diffusion and drift terms in the RG Hamiltonian, representing the non-perturbative behavior of the $O(N)$ vector model. 
Consequently, these scale-dependent propagators determine the evolution of the theory along the radial dimension.

The evolution of the theory along the radial direction $z$ is described by the change in the probability distribution $P[\varphi(x, z), \phi(x, z)]$ of the fields. 
Using the RG Hamiltonian $\mathcal{H}_{RG}$ derived in Eq. (\ref{RG_Hamiltonian}), the functional differential equation is expressed as:
\bqa && \frac{\partial}{\partial z} P[\varphi(x,z),\phi(x,z)] = N \int d^{d} x \Bigg\{ - \beta \frac{\delta}{\delta \varphi(x,z)} D(x,z) \frac{\delta}{\delta \varphi(x,z)} + \frac{\delta}{\delta \varphi(x,z)} \varphi(x,z) \nn && - \alpha \frac{\delta}{\delta \phi(x,z)} G(x,z) \frac{\delta}{\delta \phi(x,z)} + \frac{\delta}{\delta \phi(x,z)} \phi(x,z) \Bigg\} P[\varphi(x,z),\phi(x,z)] . \label{FRG_B} \eqa
Equation (\ref{FRG_B}) represents a differential equation in the functional space, describing the distribution of the field configurations as the RG scale $z$ increases. 
The diffusion terms, containing second-order functional derivatives, represent the integration of high-energy fluctuations, whereas the drift terms represent the rescaling of the fields. 

In the context of the holographic duality, this equation corresponds to a Schrödinger-like equation in the bulk, where the RG scale $z$ acts as the evolution parameter. 
The path-integral representation of this evolution yields the holographic effective action discussed in the main text, defining the relation for the emergent $(d+1)$-dimensional gravity.

The solution to the functional RG equation (\ref{FRG_B}) is expressed by integrating the RG Hamiltonian over the radial dimension. 
The probability distribution at a given scale $z$ is related to the initial distribution at $z=0$ through the evolution operator:
\bqa && P[\varphi(x,z),\phi(x,z)] = \int D \varphi(x,0) D \phi(x,0) \exp\Big\{ - \int_{0}^{z} d w \mathcal{H}_{RG} \Big\} P[\varphi(x,0),\phi(x,0)] . \hspace{0.7em}\eqa
The RG Hamiltonian is written in terms of the fields and their conjugate momenta, $\Pi_{\varphi}$ and $\Pi_{\phi}$. 
By performing a functional Fourier transform on the operators in Eq. (\ref{RG_Hamiltonian}), the RG Hamiltonian takes the canonical form:
\bqa &&\begin{split} \mathcal{H}_{RG} = N \int d^{d} x \Bigg\{ &- \beta \Pi_{\varphi}(x,w) D(x,w) \Pi_{\varphi}(x,w) + \Pi_{\varphi}(x,w) \varphi(x,w) \\
&- \alpha \Pi_{\phi}(x,w) G(x,w) \Pi_{\phi}(x,w) + \Pi_{\phi}(x,w) \phi(x,w) \Bigg\} . \end{split} \label{RG_Hamiltonian_Path_Integral} \eqa
In this representation, the terms quadratic in the momenta describe the fluctuations associated with the RG flow, while the linear terms ($\Pi_{\phi} \phi, \Pi_{\varphi} \varphi$) represent the drift. 
The initial state of the system at the UV boundary ($z=0$) is specified by the probability distribution derived from the $O(N)$ vector model:
\bqa && P[\varphi(x,0),\phi(x,0)] = \exp\Big[ - N \int d^{d} x \Big\{ \phi(x,0) G^{-1}(x,0) \phi(x,0) + \frac{1}{2 u} \varphi^{2}(x,0) \Big\} \Big] . \label{Initial_Condition} \eqa
Equation (\ref{Initial_Condition}) sets the boundary condition for the evolution. 
Propagating this UV distribution according to the Hamiltonian in Eq. (\ref{RG_Hamiltonian_Path_Integral}) yields the bulk effective action and the emergent geometry, relating the $(d+1)$-dimensional theory to the physics of the $O(N)$ vector model.


\begin{thebibliography}{99}

\bibitem{Maldacena:1997re}
J. M. Maldacena, \textit{The Large N limit of superconformal field theories and supergravity}, Adv. Theor. Math. Phys. \textbf{2}, 231 (1998) [arXiv:hep-th/9711200 [hep-th]].

\bibitem{Gubser:1998bc}
S. S. Gubser, I. R. Klebanov, and A. M. Polyakov, \textit{Gauge theory correlators from non-critical string partition functions}, Phys. Lett. B \textbf{428}, 105 (1998) [arXiv:hep-th/9802109 [hep-th]].

\bibitem{Witten:1998qj}
E. Witten, \textit{Anti-de Sitter space and holography}, Adv. Theor. Math. Phys. \textbf{2}, 253 (1998) [arXiv:hep-th/9802150 [hep-th]].

\bibitem{Aharony:1999ti}
O. Aharony, S. S. Gubser, J. Maldacena, H. Ooguri, and Y. Oz, \textit{Large N field theories, string theory and gravity}, Phys. Rept. \textbf{323}, 183 (2000) [arXiv:hep-th/9905111 [hep-th]].

\bibitem{deBoer:1999tgo}
J. de Boer, E. P. Verlinde, and H. L. Verlinde, \textit{On the holographic renormalization group}, J. High Energy Phys. \textbf{2000(08)}, 003 (2000) [arXiv:hep-th/9912012 [hep-th]].

\bibitem{Heemskerk:2009pn}
I. Heemskerk, J. Penedones, J. Polchinski, and J. Sully, \textit{Holography from Conformal Field Theory}, J. High Energy Phys. \textbf{2009(10)}, 079 (2009) [arXiv:0907.0151 [hep-th]].

\bibitem{Wetterich:1992yh}
C. Wetterich, \textit{Exact evolution equation for the effective potential}, Phys. Lett. B \textbf{301}, 90 (1993) [arXiv:1710.05815 [hep-th]].

\bibitem{Morris:1993qb}
T. R. Morris, \textit{The Exact renormalization group and approximate solutions}, Int. J. Mod. Phys. A \textbf{9}, 2411 (1994) [arXiv:hep-ph/9308265 [hep-th]].

\bibitem{Faulkner:2010jy}
T. Faulkner, H. Liu, and M. Rangamani, \textit{Integrating Out Geometry: Holographic Wilsonian RG and its Applications}, J. High Energy Phys. \textbf{2011(08)}, 051 (2011) [arXiv:1010.4036 [hep-th]].

\bibitem{Papadimitriou:2011qb}
I. Papadimitriou, \textit{Holographic Renormalization of General Dilaton-Axion Gravity}, J. High Energy Phys. \textbf{2011(08)}, 119 (2011) [arXiv:1106.4826 [hep-th]].

\bibitem{RG_Monotonicity_NEQ} 
K.-S Kim and Shinsei Ryu, \textit{Nonequilibrium thermodynamics perspectives for the monotonicity of the renormalization group flow}, Phys. Rev. D \textbf{108}, 126022 (2023) [arXiv:2310.15763 [hep-th]].

\bibitem{RG_Flow_Nonperturbative_String} 
K.-S Kim, A. Mitra, D. Mukherjee, and Shinsei Ryu, \textit{Monotonicity of the RG flow in an emergent dual holography of a world sheet nonlinear $\sigma$ model}, Phys. Rev. D \textbf{111}, 086021 (2025) [arXiv:2404.09122 [hep-th]].

\bibitem{Kim:2025frg} 
K.-S Kim, A. Mitra, D. Mukherjee, and Seung-Jong Yoo, \textit{Dual holography as functional renormalization group}, [arXiv:2511.05786 [hep-th]].

\bibitem{Klebanov:2002ja}
I. R. Klebanov and A. M. Polyakov, \textit{AdS Dual of the Critical $O(N)$ Vector Model}, Phys. Lett. B \textbf{550}, 213 (2002) [arXiv:hep-th/0210114]

\bibitem{Sathiapalan:2014mna}
R. G. Leigh, O. Parrikar, A. B. Weiss, \textit{The Exact Renormalization Group and Higher-spin Holography}, Nucl. Phys. B \textbf{894}, 183 (2015) [arXiv:1407.4574 [hep-th]].

\bibitem{Heemskerk:2010hk}
I.~Heemskerk and J.~Polchinski, \textit{Holographic and Wilsonian Renormalization Groups}, JHEP \textbf{06}, 031 (2011) [arXiv:1010.1264 [hep-th]]

\bibitem{Sathiapalan:2020pzv}
B. Sathiapalan and H. Sonoda, \textit{Holographic RG and exact RG in O(N) model}, Phys. Rev. D \textbf{91}, 026002 (2015) [arXiv:2005.10412 [hep-th]].

\bibitem{Sathiapalan:2023fuh}
P. Dharanipragada and B. Sathiapalan, \textit{Holographic RG from ERG: Locality and General Coordinate Invariance in the Bulk}, Phys. Rev. D \textbf{109}, 106017 (2024) [arXiv:2311.14446 [hep-th]].

\bibitem{Das:2003vw}
S. R. Das and A. Jevicki, \textit{Large-$N$ Collective Fields and Holography}, Phys. Rev. D \textbf{68}, 044011 (2003) [arXiv:hep-th/0304093].

\bibitem{Jevicki:2023fzp}
A. Jevicki and J. Yoon, \textit{Gravitational dynamics from collective field theory}, J. High Energy Phys. \textbf{2023(10)}, 151 (2023) [arXiv:2307.03191 [hep-th]].

\bibitem{Lee:2013dua}
S.-S. Lee, \textit{Quantum Renormalization Group and Holography}, J. High Energy Phys. \textbf{01}  (2014) 076 [arXiv:1305.3908 [hep-th]].

\bibitem{SungSik_Holography_IV} 
S.-S Lee, \textit{Horizon as Critical Phenomenon}, J. High Energy Phys. \textbf{09} (2016) 044 [arXiv:1603.08509 [hep-th]].

\bibitem{Swingle:2009bg}
B. Swingle, \textit{Entanglement Renormalization and Holography}, Phys. Rev. D \textbf{86}, 065007 (2012) [arXiv:0905.1317 [hep-th]].

\bibitem{Almheiri:2014lwa}
A. Almheiri, X. Dong, and D. Harlow, \textit{Bulk Locality and Quantum Error Correction in AdS/CFT}, J. High Energy Phys. \textbf{2015(04)}, 163 (2015) [arXiv:1411.7041 [hep-th]].

\bibitem{Pastawski:2015qua}
F. Pastawski, B. Yoshida, D. Harlow, and J. Preskill, \textit{Holographic quantum error-correcting codes: Toy models for the bulk/boundary correspondence}, J. High Energy Phys. \textbf{2015(06)}, 149 (2015) [arXiv:1503.06237 [hep-th]].

\bibitem{Hayden:2016cgc}
P. Hayden, S. Nezami, X.-L. Qi, N. Thomas, M. Walter, and Z. Yang, \textit{Holographic duality from random tensor networks}, J. High Energy Phys. \textbf{2016(11)}, 009 (2016) [arXiv:1601.01694 [hep-th]].

\bibitem{Ryu:2006bv}
S. Ryu and T. Takayanagi, \textit{Holographic Derivation of Entanglement Entropy from AdS/CFT}, Phys. Rev. Lett. \textbf{96}, 181602 (2006) [arXiv:hep-th/0603001 [hep-th]].

\bibitem{Bao:2015uaa}
N. Bao, C. Cao, S. M. Carroll, A. Chatwin-Davies, N. Hunter-Jones, J. Pollack, and G. N. Remmen, \textit{Consistency conditions for an AdS/CFT dual from entanglement}, Phys. Rev. D \textbf{91}, 125036 (2015) [arXiv:1504.06632 [hep-th]].

\bibitem{HaoGeng1:2025}
Hao Geng, Ling-Yan Hung, and Yikun Jiang, \textit{It from ETH: Multi-interval Entanglement and Replica Wormholes from Large-c BCFT Ensemble}, [arXiv:2505.20385v2 [hep-th]].

\bibitem{HaoGeng2:2025}
N. Bao, H. Geng, and Y. Jiang, \textit{Ryu-Takayanagi Formula for Multi-Boundary Black Holes from 2D Large-c CFT Ensemble}, J. High Energy Phys. \textbf{2025(10)}, 042 (2025) [arXiv:2504.12388v4 [hep-th]].

\bibitem{Czech:2015xna}
B. Czech, L. Lampros, S. McCandlish, and J. Sully, \textit{Integral Geometry and Holography}, J. High Energy Phys. \textbf{2015(10)}, 175 (2015) [arXiv:1505.05515 [hep-th]].

\bibitem{Bao:2017guc}
N. Bao, C. Cao, S. M. Carroll, and L. McInnes, \textit{Local U(1) Symmetries and the Kinematic Limit of Emergent Spacetime}, Phys. Rev. D \textbf{97}, 126009 (2018) [arXiv:1709.02399 [hep-th]].

\bibitem{Hertz:1976zz}
J. A. Hertz, \textit{Quantum critical phenomena}, Phys. Rev. B \textbf{14}, 1165 (1976).

\bibitem{Millis:1993zz}
A. J. Millis, \textit{Effect of a nonzero temperature on quantum critical points in itinerant fermion systems}, Phys. Rev. B \textbf{48}, 7183 (1993).

\bibitem{Metzner:2011zz}
W. Metzner, M. Salmhofer, C. Honerkamp, V. Meden, and K. Schönhammer, \textit{Functional renormalization group approach to correlated electron systems}, Rev. Mod. Phys. \textbf{84}, 299 (2012) [arXiv:1110.1457 [cond-mat.str-el]].

\bibitem{SungSik_Holography_III} 
Peter Lunts, Subhro Bhattacharjee, Jonah Miller, Erik Schnetter, Yong Baek Kim, and Sung-Sik Lee, \textit{Ab initio holography}, J. High Energy Phys. \textbf{08} (2015) 107 [arXiv:1503.06474 [hep-th]].

\bibitem{Koch:2010cy}
R. de Mello Koch, A. Jevicki, K. Jin, and J. P. Rodrigues, \textit{AdS4/CFT3 Construction from Collective Fields}, Phys. Rev. D \textbf{83}, 025006 (2011) [arXiv:1008.0633 [hep-th]].

\bibitem{Brute_Force_RG_Derivation_Lattice_Kim} 
K.-S. Kim and C. Park, \textit{Emergent geometry from field theory: Wilson's renormalization group revisited}, Phys. Rev. D \textbf{93}, 121702 (2016) [arXiv:1604.04990 [hep-th]].

\bibitem{Kitaev_Entanglement_Entropy_Kim} 
K.-S. Kim, M. Park, J. Cho, and C. Park, \textit{An emergent geometric description for a topological phase transition in the Kitaev superconductor model}, Phys. Rev. D \textbf{96}, 086015 (2017) [arXiv:1610.07312 [hep-th]].
	
\bibitem{Kondo_Holography} 
K.-S Kim, Suk Bum Chung, Chanyong Park, and Jae-Ho Han, \textit{A non-perturbative field theory approach for the Kondo effect: Emergence of an extra dimension and its implication for the holographic duality conjecture}, Phys. Rev. D \textbf{99}, 105012 (2019) [arXiv:1705.06571 [hep-th]].			
\bibitem{RG_GR_Geometry_I_Kim} 
K.-S Kim, \textit{Geometric encoding of renormalization group $\beta$-functions in an emergent holographic dual description}, Phys. Rev. D \textbf{102}, 026022 (2020) [arXiv:2003.00281 [cond-mat.str-el]].

\bibitem{RG_GR_Geometry_II_Kim} 
K.-S Kim, \textit{Emergent geometry in recursive renormalization group transformations}, Nucl. Phys. B \textbf{959}, 115144 (2020) [arXiv:2004.09997 [cond-mat.str-el]].			

\bibitem{Brute_Force_RG_Derivation_Dirac_Kim} 
K.-S Kim, \textit{Emergent dual holographic description for interacting Dirac fermions in the large $N$ limit}, Phys. Rev. D \textbf{102}, 086014 (2020) [arXiv:2003.00291 [cond-mat.str-el]].	
	
\bibitem{RG_Flow_Direct_Calculation} 
K.-S Kim and Shinsei Ryu, \textit{Entanglement transfer from quantum matter to classical geometry in an emergent holographic dual description of a scalar field theory}, J. High Energy Phys. \textbf{05} (2021) 260, [arXiv:2003.00165 [hep-th]].

\bibitem{Nonperturbative_Wilson_RG} 
K.-S Kim, Shinsei Ryu, and Kanghoon Lee, \textit{Emergent dual holographic description as a nonperturbative generalization of the Wilsonian renormalization group}, Phys. Rev. D \textbf{105}, 086019 (2022) [arXiv:2112.06237 [hep-th]].

\bibitem{Nonperturbative_Wilson_RG_Disorder} 
K.-S Kim, \textit{Beyond quantum chaos in emergent dual holography}, Phys. Rev. D \textbf{106}, 126014 (2022) [arXiv:2210.17192 [cond-mat.str-el]].

\bibitem{Nonperturbative_RG_Flow} 
K.-S Kim, Mitsuhiro Nishida, and Yoonseok Choun, \textit{Renormalization group flow to effective quantum mechanics at IR in an emergent dual holographic description for spontaneous chiral symmetry breaking}, Phys. Rev. D \textbf{107}, 066004 (2023) [arXiv:2210.10277 [hep-th]].

\bibitem{Abramowitz:1972}
M. Abramowitz and I. A. Stegun, \textit{Handbook of Mathematical Functions with Formulas, Graphs, and Mathematical Tables}, Dover (1972).

\bibitem{Olver:2010}
F. W. J. Olver, D. W. Lozier, R. F. Boisvert, and C. W. Clark, \textit{NIST Handbook of Mathematical Functions}, Cambridge University Press (2010).

\bibitem{Jost:2008}
J. Jost, \textit{Riemannian Geometry and Geometric Analysis}, Springer (2008).

\bibitem{Polchinski:1998rq}
J. Polchinski, \textit{String Theory. Vol. 1: An Introduction to the Bosonic String}, Cambridge University Press (1998).








%
%
%
%
%
%
%




\bibitem{Kontou:2020bta}
E.-A.~Kontou and K.~Sanders, \textit{Energy Conditions in General Relativity and Quantum Field Theory}, Class.\ Quant. \ Grav.\ \textbf{37}, no.~19, 193001 (2020) [arXiv:2003.01815 [gr-qc]]

\bibitem{Fitzpatrick:2014vua}
A. Liam Fitzpatrick, Jared Kaplan, and Matthew T. Walters,
\textit{Universality of Long-Distance AdS Physics from the CFT Bootstrap}, JHEP \textbf{08}, 145 (2014) [arXiv:1403.6829 [hep-th]].

\end{thebibliography}
\end{document}